\documentclass[a4paper,fleqn]{cas-dc}

\usepackage[numbers,sort&compress]{natbib}

\bibpunct{(}{)}{,}{a}{,}{,}
\def\tsc#1{\csdef{#1}{\textsc{\lowercase{#1}}\xspace}}
\tsc{WGM}
\tsc{QE}

\usepackage{booktabs}
\usepackage{multirow}
\usepackage{graphicx}
\usepackage{placeins}
\usepackage{algpseudocode}
\usepackage[table]{xcolor}
\usepackage{xcolor}
\usepackage{bm}
\usepackage{silence}
\usepackage{threeparttable}
\usepackage{subcaption}
\usepackage{siunitx}
\usepackage{afterpage}
\usepackage{xcolor}

\usepackage{soul}

\ifdefined\pdfsuppresswarningpagegroup
\fi

\makeatletter
\let\font@warning\@gobble
\makeatother
\hypersetup{hypertexnames=false}

\ExplSyntaxOn
\RenewDocumentEnvironment { Abstract } { o }
{
  \group_begin:
  \IfNoValueTF { #1 } { }
    { \tex_gdef:D \abstractname { #1 } }
  \parindent 0pt
  \box_if_empty:NTF \g_stm_key_box
    { \leftskip = .35 \textwidth }
    {
      \dim_gset:Nn \l_tmpa_dim { \box_ht:N \g_stm_key_box }
      \dim_gadd:Nn \l_tmpa_dim { \box_dp:N \g_stm_key_box }
      \leftskip .35\textwidth
      \hspace*{-.35 \textwidth }
      \noindent\hbox_to_wd:nn { 0pt } { \box \g_stm_key_box \hss }
      \skip_vertical:n { - \l_tmpa_dim }
    }
  \noindent \abstractname \par
  \skip_vertical:n { -4pt }
  \noindent \rule{.65\textwidth}{.2pt}\par \footnotesize
  \ignorespaces \everypar { \parindent=1.5em }
}
{ \par \group_end: }
\ExplSyntaxOff

\definecolor{JournalBlue}{HTML}{1F4E79}
\definecolor{JournalLightBlue}{HTML}{EAF1F7}
\definecolor{bestblue}{HTML}{E7F0F7}
\definecolor{secondred}{HTML}{F7E3E5}
\definecolor{improvegreen}{HTML}{E8F2EC}
\definecolor{regressred}{HTML}{F7E3E5}
\definecolor{neutralgray}{HTML}{F3F3F3}
\definecolor{deltaup}{HTML}{478B68}
\definecolor{deltadown}{HTML}{B56B75}
\definecolor{rankednavy}{HTML}{17365D}
\definecolor{biasgood}{HTML}{E8F2EC}
\definecolor{biasbad}{HTML}{F7E3E5}
\makeatletter
\def\InlineResultBreak{\@ifnextchar[{\InlineResultBreak@i}{\nobreak\hspace{0.18em}}}
\def\InlineResultBreak@i[#1]{\nobreak\hspace{0.18em}}
\newcommand{\InlineResult}[1]{%
  \begingroup
  \let\\\InlineResultBreak
  #1%
  \endgroup}

\newcommand{\CompactDeltaResult}[2]{%
  \begingroup
  \setbox\z@=\hbox{#1}%
  \setbox\tw@=\hbox{{\fontsize{5.0}{5.0}\selectfont #2}}%
  \dimen@=\wd\z@
  \ifdim\wd\tw@>\dimen@ \dimen@=\wd\tw@\fi
  \vtop{%
    \offinterlineskip
    \hbox to \dimen@{\hfil\box\z@\hfil}%
    \vskip0.3pt
    \hbox to \dimen@{\hfil\box\tw@\hfil}}%
  \endgroup}
\makeatother
\algrenewcommand\algorithmicrequire{\textbf{Input:}}
\algrenewcommand\algorithmicensure{\textbf{Output:}}

\newcommand{\ourmethod}{DAMM-Net++}

\newcommand{\ours}{DAMM-Net++}

\renewcommand{\figurename}{Fig.}

\DeclareCaptionLabelFormat{unnumberedalgorithm}{Algorithm}

\begin{document}
\let\WriteBookmarks\relax
\setcounter{topnumber}{4}
\setcounter{bottomnumber}{2}
\setcounter{totalnumber}{6}
\setcounter{dbltopnumber}{4}
\renewcommand{\topfraction}{0.95}
\renewcommand{\bottomfraction}{0.85}
\renewcommand{\textfraction}{0.05}
\renewcommand{\floatpagefraction}{0.70}
\renewcommand{\dbltopfraction}{0.95}
\renewcommand{\dblfloatpagefraction}{0.60}
\hbadness=10000
\hfuzz=0.5pt
\emergencystretch=2em

\shorttitle{\hspace{0pt}}    

\shortauthors{\hspace{0pt}}  

\title [mode = title]{Anatomy-Change-Aware Bidirectional Selective State-Space Memory for Clinically Deployed Thoracic Radiotherapy Auto-Contouring} 




\author[2,3]{Galib Ahmed}
\fnmark[\noexpand\textdagger]
\ead{galib.ahmed.251@northsouth.edu}
\credit{Conceptualization, Data Curation, Formal analysis, Investigation, Methodology, Software, Validation, Visualization, Writing – original draft, Writing – review and editing}

\author[1,3]{Istiak Ahmed}
\fnmark[\noexpand\textdagger]
\ead{istiak.ahmed1@northsouth.edu}
\credit{Conceptualization, Data Curation, Formal analysis, Investigation, Methodology, Software, Validation, Visualization, Writing – original draft, Writing – review and editing}

\author[1,3]{Aritra Islam Saswato}
\ead{aritra.saswato@northsouth.edu}
\credit{Data Curation, Formal Analysis, Methodology, Software, Validation, Visualization, Writing – review and editing}

\author[3,9]{Asib Mostakim Fony}
\ead{asibmostakim@ius.edu.bd}
\credit{Data Curation, Formal Analysis, Methodology, Software, Validation, Visualization, Writing – review and editing}

\author[1,3]{Kazi Shahriar Sanjid}
\ead{kazi.sanjid@northsouth.edu}
\credit{Data Curation, Formal Analysis, Methodology, Software, Validation, Visualization, Writing – review and editing}

\author[3,4]{Md. Tanzim Hossain}

\ead{tanzim.hossain@fau.de}
\credit{Formal Analysis, Software, Validation, Visualization, Writing – review and editing}

\author[3,5]{Md. Anwarul Islam}
\ead{anwarpabna@gmail.com}
\credit{Data Curation, Formal Analysis, Resources, Validation, Writing – review and editing}

\author[1,3]{Md. Nishan Khan}
\ead{nishan.khan@northsouth.edu}
\credit{Data Curation, Methodology, Software, Validation}

\author[1,3]{Md. Misbah Khan}
\ead{misbah.khan@northsouth.edu}
\credit{Data Curation, Methodology, Software, Visualization}

\author[1,3]{Labiba Faiza Karim}
\ead{labiba.karim@northsouth.edu}
\credit{Data Curation, Formal Analysis, Methodology, Visualization, Writing – review and editing}

\author[11]{Jobaer Rahman}
\ead{mra002@email.latech.edu}
\credit{Data Curation, Software, Validation, Visualization, Writing – review and editing}

\author[6]{S M Hasibul Hoque}
\ead{hasibulshefat@gmail.com}
\credit{Conceptualization, Resources, Validation, Writing – review and editing}

\author[2]{Rahnuma Shahrin Rista}
\ead{rahnuma.rista@northsouth.edu}
\credit{Conceptualization, Resources, Validation, Writing – review and editing}

\author[3,9]{Kamruzzaman Rumman}
\ead{drarifur@bshl.com.bd}
\credit{Formal Analysis, Resources, Validation, Visualization}

\author[7]{Md Arifur Rahman}
\ead{drarifur@bshl.com.bd}
\credit{Formal Analysis, Resources, Validation, Visualization}

\author[8]{Syed Md. Akram Hussain}
\ead{syedmdakram@gmail.com}

\credit{Formal Analysis, Resources, Project Administration, Validation, Writing – review and editing}

\author[1,3]{Mohammad Ashrafuzzaman Khan}
\ead{mohammad.khan02@northsouth.edu}
\credit{Data Curation, Formal Analysis, Methodology, Software, Visualization}

\author[2,3]{M. Monir Uddin}
\cormark[1]
\ead{monir.uddin@northsouth.edu}
\credit{Conceptualization, Formal Analysis, Funding acquisition, Methodology, Project administration, Resources, Supervision, Writing – review and editing}

\nonumnote{\textsuperscript{\textdagger} These authors contributed equally to this work.}


\affiliation[1]{%
        addressline={Department of Electrical and Computer Engineering},
        organization={North South University},
    city={Dhaka},
    postcode={1229},
    country={Bangladesh}
}

\affiliation[2]{%
        addressline={Department of Mathematics and Physics},
        organization={North South University},
    city={Dhaka},
    postcode={1229},
    country={Bangladesh}
}

\affiliation[3]{%
addressline={Big-Matrix Lab, Department of Mathematics and Physics},
    organization={North South University},
   city={Dhaka},
    postcode={1229},
    country={Bangladesh}
}

\affiliation[4]{%
addressline={Department of Data Science},
    city={Erlangen},
    organization={Friedrich-Alexander University},
        postcode={91054},
    country={Germany}
}

\affiliation[5]{%
    addressline={Square Cancer Centre},
    organization={Square Hospitals Limited},
       city={},
    postcode={Dhaka 1205},
    country={Bangladesh}
}

\affiliation[6]{%
addressline={Department of Radiation Oncology},
    organization={Labaid Cancer Hospital \& Super Speciality Center},
        city={},
    postcode={Dhaka 1205},
    country={Bangladesh}
}

\affiliation[7]{%
    addressline={Department of Oncology \& Radiotherapy},
    organization={Bangladesh Specialized Hospital Limited},
       city={},
    postcode={Dhaka 1207},
    country={Bangladesh}
}

\affiliation[8]{%
    addressline={Department of Clinical Oncology},
    city={},
    organization={Bangladesh Medical University},
       postcode={Dhaka 1000},
    country={Bangladesh}
}

\affiliation[9]{%
    addressline={Department of Radiation Oncology},
    city={Dhaka},
    organization={National Institute of Cancer Research \& Hospital},
       postcode={1212},
    country={Bangladesh}
}

\affiliation[10]{%
    addressline={Department of Computer Science \& Engineering},
    city={Dhaka},
    organization={University of Scholars},
       postcode={1213},
    country={Bangladesh}
}

\affiliation[11]{%
    addressline={Department of Computer Science \& Engineering},
    city={},
    organization={Louisiana Tech University},
       postcode={201 Mayfield Ave, Ruston, LA 71272},
    country={USA}
}

\cortext[1]{Corresponding author}





\begin{abstract}
We developed \ours{}, a 2.5D architecture for thoracic OAR and target volume segmentation that addresses three persistent challenges in radiotherapy auto-contouring: inter-slice surface incoherence, systematic failure on small low-contrast targets, and the absence of per-case reliability signals. The central component is an anatomy-change-aware bidirectional selective state-space memory that models through-plane anatomical change and selectively propagates context along the axial slice sequence. A boundary-aware decoder sharpens near-surface predictions, and an uncertainty head provides calibrated per-voxel confidence for clinical triage. We evaluated 2,146 patients across four centers, an independent external cohort of 112 patients, and a multicenter reader study involving 17 radiation oncologists on 305 cases. The model achieves a mean Dice of 0.955 and HD95 of 3.78 mm, with the largest gains on low-contrast organs-at-risk (OARs) and target volumes where through-plane context is most critical. The uncertainty head is well-calibrated and supports case-level triage. In the reader study, AI assistance reduced contouring time by 75--80\% across experience levels and raised junior-reader IoU from 0.861 to 0.925, matching the unedited model. External validation showed a modest internal-to-external drop ($<5\%$) with calibrated uncertainty transferring without recalibration. The complete deployment pipeline from DICOM ingestion to TPS-compatible RTSTRUCT export has been integrated into the clinical workflow at a partner hospital, where it is used to assist with contouring. These results suggest that anatomically motivated inter-slice memory, paired with uncertainty-guided review, offers a clinically viable path for thoracic auto-contouring.
\end{abstract}
\begin{keywords}
Thoracic radiotherapy \sep Radiotherapy auto-contouring \sep Anatomy-change-aware memory \sep Bidirectional state-space model \sep Multi-center reader study \sep Uncertainty-guided triage \sep Clinical deployment
\end{keywords}

\begin{NoHyper}
\maketitle
\end{NoHyper}





\section{Introduction}
\label{sec:introduction}

Lung cancer remains the leading cause of cancer-related mortality worldwide, accounting for approximately 1.9 million deaths in 2024 \citep{sung2021global}. Radiation therapy is a cornerstone of treatment for many patients with lung cancer \citep{delaney2005radiotherapy}, and its effectiveness depends critically on accurate delineation of organs-at-risk (OARs) and target volumes on planning CT. Contouring errors can affect dose distribution, tumor control, and toxicity risk \citep{marks2010radiation, bentzen2010normal}. However, manual contouring is time-intensive, operator-dependent, and subject to inter-observer variability \citep{ref_gtv_interobserver, Vinod2016}. These challenges are particularly consequential in settings with limited radiotherapy workforce capacity \citep{ref_bangladesh_workforce} and unequal access to radiotherapy services \citep{ref_lmic_radiotherapy_gap}.

Deep learning has substantially advanced automated medical image segmentation \citep{Ronneberger2015, ref_3dunet, ref_attention_unet, ref_transunet, ref_swin, Gu2023}, yet thoracic radiotherapy auto-contouring remains challenging because anatomy evolves continuously along the craniocaudal axis while varying substantially in shape, size, and contrast. Two-dimensional approaches lack explicit inter-slice context, whereas full 3D architectures impose substantial computational and memory requirements \citep{ref_nnunet, ref_unetr}. Sequence-based 2.5D methods provide an intermediate alternative, but conventional recurrent formulations such as ConvLSTM and ConvGRU employ fixed transition mechanisms that do not adapt to the anatomical content of individual slice-to-slice transitions \citep{novikov2018deepsequential, shi2015convlstm}.

This limitation is particularly important for structures whose appearance changes rapidly or irregularly across adjacent slices. The tapering trachea, shifting mediastinal anatomy, and appearance or disappearance of target volumes require selective retention of contextual information rather than uniform propagation. A related challenge is the segmentation of GTV and CTV, which is affected by foreground-background imbalance, morphological variability, and limited intrinsic contrast \citep{ref_ctv_autoseg_review_2024, ref_lung_gtv_deep}.

We therefore hypothesize that thoracic auto-contouring can benefit from an inter-slice memory whose state transition is conditioned on the observed anatomical change between adjacent slices. We instantiate this in \ours{} (Dynamic Anatomical Memory Network), which introduces an anatomy-change-aware bidirectional selective state-space memory into a 2.5D segmentation framework. A Mamba-based state-space model uses input-dependent dynamics conditioned on a multi-scale representation of inter-slice anatomical change, allowing information to be selectively retained or suppressed as anatomy evolves along the scan. Bidirectional processing integrates context from both superior-to-inferior and inferior-to-superior directions. Around this core, \ours{} incorporates a memory-guided boundary-aware decoder for boundary refinement and an uncertainty head for predictive confidence. Auxiliary objectives, including signed-distance regression, organ-presence classification, and topology regularization, provide complementary supervision.

We test the central hypothesis using controlled transition experiments that compare real inter-slice anatomical transitions with randomized and shuffled controls, thereby assessing whether anatomical-change information itself contributes to segmentation performance. We further evaluate uncertainty calibration, external generalization, reader performance, and clinical workflow integration.

We evaluate \ours{} on a multi-institutional dataset of 2,146 patients across four Bangladeshi tertiary centers, encompassing nine thoracic structures. We benchmark against nnUNet \citep{ref_nnunet}, nnMamba \citep{gong2025nnmamba}, UNETR++ \citep{shaker2024unetrpp}, and SwinUNETR \citep{hatamizadeh2022aswinunetr}. Beyond retrospective benchmarking, we present a multicenter reader study with 17 oncologists, external validation on 112 held-out patients, uncertainty analysis, and an end-to-end clinical deployment pipeline from DICOM ingestion to TPS-compatible RTSTRUCT export at Bangladesh Medical University.

Our contributions are:
\begin{itemize}
\item An anatomy-change-aware bidirectional selective state-space memory that conditions state-space dynamics on observed inter-slice anatomical transitions within a 2.5D thoracic CT framework.
\item A controlled ablation demonstrating that real anatomical-transition information, rather than arbitrary feature augmentation, drives the accuracy gains on CTV, GTV, esophagus, and trachea.
\item A memory-guided boundary-aware decoder and single-pass uncertainty head for boundary refinement and case-level predictive quality assessment.
\item Multicenter reader evaluation with 17 radiation oncologists demonstrating 75--80\% reductions in contouring time and an increase in junior-reader IoU from 0.861 to 0.925.
\item An end-to-end clinical pipeline from DICOM ingestion to TPS-compatible RTSTRUCT export, evaluated within a clinician-in-the-loop workflow.
\end{itemize}

\section{Related work}
\label{sec:related_work}
We organize this review around the four gaps identified in the introduction, focusing on the methodological strands most relevant to our contributions.

\subsection{Thoracic radiotherapy auto-contouring}

Inter-observer variability remains a persistent source of uncertainty in radiotherapy contouring, directly affecting target coverage and OAR sparing \citep{Vinod2016, Nelms2012}. A recent systematic review found that auto-contouring generally improves inter-observer consistency, although benefits vary by structure and clinician review remains necessary \citep{darby2026interobserver}. Early deep-learning methods demonstrated that individual OAR delineation could approach expert-level accuracy \citep{Vaassen2020, vandijk2020, Cardenas2018, Cardenas2019, Zhu2020, vanRooij2019}, and the self-configuring nnU-Net \citep{ref_nnunet} became a widely adopted baseline.

Two structural challenges persist. First, target volumes (GTV, CTV) show systematically larger boundary errors than OARs, reflecting both ambiguity at tumor-tissue interfaces and inter-observer disagreement \citep{ref_gtv_interobserver, ref_ctv_autoseg_review_2024}. Second, the esophagus remains challenging on thoracic CT because of low contrast and anatomical variability \citep{ref_rt_thoracic_oar}, and a recent multicenter study reported lower accuracy for trachea and esophagus in patients with lung cancer than in those without \citep{niu2026thoracic}. These are precisely the structures on which our method shows the largest gains over the volumetric baselines.

Recent work has begun to address clinical ambiguity and prospective deployment more directly. Balagopal et al. demonstrated uncertainty-aware segmentation of anatomically invisible postoperative prostate CTVs, providing a precedent for ambiguity-aware target delineation \citep{balagopal2021invisible}. A prospective five-center thoracic trial evaluated AI-assisted OAR delineation in 500 patients with 37 physicians, reporting improved geometric agreement and substantially reduced contouring time relative to manual delineation \citep{niu2026thoracic}.

\subsection{Through-plane context modeling}

Purely 2D methods discard inter-slice information, producing label fields that oscillate across slices \citep{ref_2dvs3d}, while volumetric 3D methods integrate context natively but at substantial memory cost from cropped or downsampled patches \citep{Alalwan2021}. Representative volumetric approaches include nnU-Net, Swin UNETR, UNETR++, and nnMamba \citep{ref_nnunet, hatamizadeh2022aswinunetr, shaker2024unetrpp, gong2025nnmamba}, where GPU memory is coupled with patch and batch sizes.

Between these extremes, 2.5D designs process adjacent slice sequences using recurrent inter-slice memory such as ConvLSTM and ConvGRU \citep{shi2015convlstm, novikov2018deepsequential}, with bidirectional processing reducing directional bias \citep{Zhang2015}. Two properties limit their radiotherapy fit: transition operators are fixed after training and cannot parameterize dynamics from each observed transition, and sequential updates complicate full-thorax training. Video propagation methods offer temporal coherence but do not integrate a selective mechanism conditioned on explicit inter-slice anatomical change, which is the central focus of our architecture.

\subsection{Selective state-space models}

Selective state-space models (SSMs), particularly Mamba \citep{Gu2023}, offer linear time complexity in sequence length, a constant-size recurrent state during inference, and input-dependent discretization enabling selective retention based on the current input. Recent work applies Mamba to medical segmentation \citep{Xing2024, Ma2024a, Wang2024a, Ruan2024, Liu2024a}, either as global-context blocks alongside convolutional encoders \citep{gong2025nnmamba} or as replacements for self-attention in transformer hybrids \citep{Xing2024, Wang2024a}.

Our design takes SSMs in a different direction. Rather than using the standard selective update inside a volumetric backbone, we make the selection signal itself anatomically motivated: the discretization parameters are conditioned on a multi-scale inter-slice transition branch that observes actual inter-slice anatomical change. The resulting selective memory is run bidirectionally along the axial slice sequence. To our knowledge, no prior medical imaging study explicitly conditions selective state-space dynamics on an inter-slice anatomical transition representation in this setting.

\subsection{Auxiliary components}

Boundary-aware supervision addresses the limitation of overlap losses that treat all foreground voxels identically, under-weighting near-surface voxels that dominate dose calculations. For clinical quality assurance, single-pass aleatoric uncertainty estimation \citep{Kendall2017} offers an efficient alternative to expensive Bayesian approximations \citep{Gal2016, Lakshminarayanan2017}. Calibration and case-level triage utility require explicit evaluation rather than being inferred from segmentation accuracy alone \citep{huang2024uncertainty, wahid2024uncertainty}, as we evaluate in \hyperref[sssec:uncertainty]{Section~\ref{sssec:uncertainty}}.

\section{Methods}
\label{sec:methods}
This section describes the multicenter thoracic radiotherapy planning dataset used for training and evaluation, the proposed DAMM-Net++ architecture and its core components, and the experimental setup used for benchmarking.

\subsection{Dataset preparation}
\label{sec:dataset}

\begin{table*}[pos=htbp]
\centering
\caption{Institutional distribution of the multicenter thoracic radiotherapy planning cohort. BMU, SHL, and LH comprised the internal cohort split 70/15/15; UHL was reserved exclusively for external validation.}
\label{tab:cohort}
\begin{tabular}{lcccc}
\toprule
\textbf{Institution} & \textbf{Patients} & \textbf{Proportion (\%)} & \textbf{Acquisition Years} & \textbf{Role} \\
\midrule
\multicolumn{5}{c}{\textbf{Internal Cohort}} \\
Bangladesh Medical University (BMU) & 1,447 & 67.4 & 2020--2026 & Internal \\
Square Hospital Limited (SHL)       & 309   & 14.4 & 2018--2026 & Internal \\
Labaid Hospital (LH)                & 278   & 13.0 & 2019--2025 & Internal \\
\midrule
\multicolumn{5}{c}{\textbf{External Validation Cohort}} \\
United Hospital (UHL)               & 112   & N/A  & 2026 & External Validation \\
\midrule
\textbf{Total}                      & \textbf{2,146} & \textbf{100.0} & N/A & N/A \\ 
\multicolumn{5}{c}{\emph{Internal split: Training $1{,}424$ (70\%) $\cdot$ Validation $305$ (15\%) $\cdot$ Test $305$ (15\%)}} \\
\bottomrule
\end{tabular}
\end{table*}

The retrospective, multi-institutional dataset comprised thoracic radiotherapy planning CTs from four Bangladeshi tertiary hospitals over approximately eight years (2018--2026), ensuring technological diversity across scanner generations (\hyperref[tab:cohort]{Table~\ref{tab:cohort}}). For each patient, the planning CT and corresponding DICOM RTSTRUCT were retrieved, encoding clinically approved delineations of nine anatomical structures. All CT volumes were reconstructed at $512 \times 512$ in-plane resolution with mean spacing $1.009 \pm 0.081$~mm and mean inter-slice spacing $2.493 \pm 0.097$~mm, ranging from 100 to over 350 slices (mean: 181). Cases with truncation, severe artifacts, or incomplete contours were excluded, yielding 2,146 patients.

RTSTRUCT contours were produced by board-certified radiation oncologists following ESTRO and RTOG guidelines, with harmonized nomenclature and geometric validation. The spinal cord, heart, lungs, and body were delineated in nearly all patients, while the esophagus and trachea showed lower occurrence rates due to institutional variability. GTV and CTV exhibited the most severe foreground-background imbalance, motivating imbalance-aware losses, as described in \hyperref[subsec:multi_task_loss]{Section~\ref{subsec:multi_task_loss}}. An inter-rater study on 100 randomly selected cases with ten oncologists independently delineating all structures showed IoU exceeding 0.95 for every structure, confirming high annotation reliability.

The internal cohort of 2,034 patients was split 70/15/15 into training ($n=1{,}424$), validation ($n=305$), and test ($n=305$) sets using institution-stratified sampling, with the 112 UHL patients withheld as an independent external validation cohort.

DICOM RTSTRUCT files were converted to volumetric segmentation masks using 3D Slicer. A canonical label mapping and priority ordering resolved overlaps, CT intensities were windowed (center: 50~HU, width: 400~HU; range $[-150, 250]$~HU), normalized to $[0, 255]$, resampled to $512 \times 512$, and decomposed into axial slices. Only axial slices were used for training and evaluation, matching the standard clinical contouring view. The complete pipeline is detailed in \hyperref[alg:dicom_pipeline]{Algorithm~\ref{alg:dicom_pipeline}}.

\subsection{Model architecture}
\label{subsec:architecture}

\begin{figure*}[pos=htbp]
    \centering
    \includegraphics[width=\textwidth]{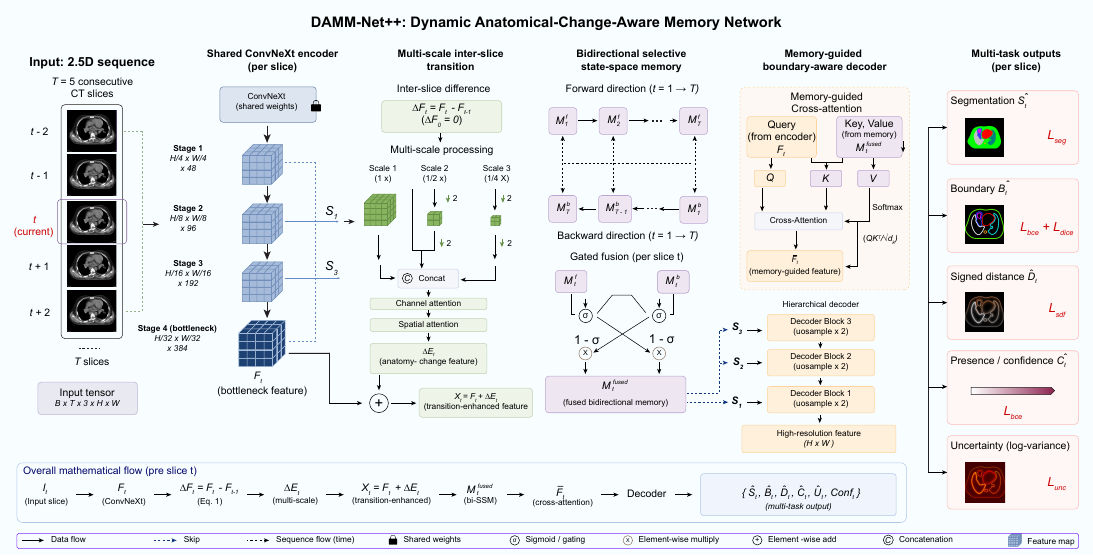}
    \caption{Overview of the proposed DAMM-Net++ architecture. The input is a sequence of $T$ consecutive axial CT slices, each passed through a weight-shared ConvNeXt encoder. The Multi-Scale Inter-Slice Transition branch computes inter-slice feature differences; the Bidirectional Selective State-Space Memory accumulates anatomy-change-informed context in both through-plane directions; and the Memory-Guided Boundary-Aware Decoder reconstructs full-resolution predictions. The network produces segmentation, boundary, signed distance map, presence, and uncertainty outputs.}
    \label{fig:model_architecture}
\end{figure*}

DAMM-Net++ is a 2.5D sequence-to-sequence segmentation network designed for thoracic OAR and target volume delineation. Rather than processing each axial slice independently, the network accepts a sequence of $T = 5$ consecutive slices as a single input tensor of shape $B \times T \times 3 \times 512 \times 512$. This multi-slice context is clinically motivated, as radiation oncologists routinely scroll through adjacent slices when contouring, using spatial continuity to resolve anatomical ambiguity and distinguish true boundaries from imaging artifacts. The network architecture comprises four principal components: shared ConvNeXt encoder, multi-scale inter-slice transition branch, bidirectional state-space memory module, and memory-guided boundary-aware decoder, illustrated in \hyperref[fig:model_architecture]{Fig.~\ref{fig:model_architecture}}. Detailed views of the transition branch, memory module, and cross-attention are provided in \hyperref[fig:detail_module]{Fig.~\ref{fig:detail_module}}.

The forward pass through \ours{} follows a cascaded transformation as

\begin{flalign}
\label{eq:forward_pass}
\begin{aligned}
    I_t \xrightarrow{\text{ConvNeXt}} F_t 
    &\xrightarrow{\text{Transition}} \Delta E_t \\
    X_t &= F_t + \Delta E_t \\
    X_t \xrightarrow{\text{Bi-SSM}} M_t 
    &\xrightarrow{\text{Gate}} M_t^{\text{fused}} \\
    \tilde{F}_t &= \text{CrossAttn}(F_t, M_t^{\text{fused}}) \\
    \tilde{F}_t, M_t^{\text{fused}} &\xrightarrow{\text{Decoder}} \hat{Y}_t \\
    \hat{Y}_t &\xrightarrow{\text{Heads}} \{\hat{S}_t, \hat{B}_t, \hat{D}_t, \hat{C}_t\}
\end{aligned} &&
\end{flalign}

where $I_t$ is the input slice, $F_t$ is the bottleneck feature, $\Delta E_t$ is the anatomy-change-enhanced feature, $M_t^{\text{fused}}$ is the fused bidirectional memory, $\hat{Y}_t$ is the decoded feature map, $\hat{S}_t$ is the segmentation, $\hat{B}_t$ is the boundary, $\hat{D}_t$ is the signed distance, and $\hat{C}_t$ is the confidence outputs.

\subsubsection{ConvNeXt encoder}

A weight-shared ConvNeXt encoder processes each of the $T$ slices independently, producing a four-level feature pyramid. The encoder uses a $4\times4$ stem followed by four stages with $(2, 2, 6, 2)$ ConvNeXt blocks per stage, where each block applies a depthwise $7\times7$ convolution, layer normalization, and an inverted bottleneck with GELU activation and stochastic depth regularization. The four stages produce feature maps at resolutions $H/4$, $H/8$, $H/16$, and $H/32$ with channel dimensions $48, 96, 192$, and $384$, respectively. The three shallower feature maps serve as skip connections to the decoder, while the deepest bottleneck feeds the inter-slice transition and memory modules. No inter-slice mixing occurs in the encoder; through-plane reasoning is introduced exclusively through the downstream transition and memory components, consistent with the overall forward pass shown in \hyperref[eq:forward_pass]{Eq.~\eqref{eq:forward_pass}}.

\subsubsection{Multi-scale inter-slice transition branch}

The inter-slice transition branch models anatomical change between adjacent slices by computing the feature-space difference at the bottleneck level a:

\begin{equation}
\label{eq:inter_slice_diff}
    \Delta F_t = F_t^{(4)} - F_{t-1}^{(4)}
\end{equation}

with $\Delta F_0 = 0$ for the first slice. To capture anatomical changes at multiple scales, the difference map $\Delta F_t$ from \hyperref[eq:inter_slice_diff]{Eq.~\eqref{eq:inter_slice_diff}} is processed at three scales native, half, and quarter resolution. The multi-scale features are then fused and refined through channel and spatial attention to produce the anatomy-change-enhanced feature $\Delta E_t$, illustrated in \hyperref[fig:detail_module]{Fig.~\ref{fig:detail_module}a}. The final transition output is added as a weighted residual to the original difference: $\Delta F_t + \gamma_m \cdot \Delta E_t$, where $\gamma_m$ is a learnable scalar initialized to zero. This design allows the network to attend to both fine-scale anatomical changes, such as small vessel appearance or boundary shifts, and coarse-scale changes, such as lung volume variation or organ entry and exit across slices.

\subsubsection{Bidirectional selective state-space memory}

The memory module accumulates anatomy-change-informed context across the slice sequence using a selective state-space model (SSM) following Mamba~\citep{Gu2023}, which dynamically adjusts its state-space dynamics according to the input representation. In our formulation, the bottleneck representation is augmented with an inter-slice anatomical transition signal before determining the SSM parameters.

Let $F_t^{(4)}$ denote the bottleneck at axial position $t$. The inter-slice transition $\Delta F_t$ from \hyperref[eq:inter_slice_diff]{Eq.~\eqref{eq:inter_slice_diff}} is transformed into an anatomy-change-enhanced representation $\Delta E_t = f_{\mathrm{trans}}(\Delta F_t)$. The resulting input to the SSM is $X_t = F_t^{(4)} + \Delta E_t$, so the transition signal directly participates in the state-space dynamics.

The SSM generates input-dependent parameters as

\begin{equation}
\label{eq:ssm_params}
\Delta_t = f_{\Delta}(X_t), \qquad
B_t = f_B(X_t), \qquad
C_t = f_C(X_t)
\end{equation}

where $\Delta_t$ controls the discretization step, while $B_t$ and $C_t$ parameterize the input and output projections. The discretized transition matrix $\bar{A}_t$ is obtained from $\Delta_t$ and the learned continuous-time matrix $A$, while $\bar{B}_t$ is obtained from $\Delta_t$ and $B_t$. The recurrent state update is $s_t = \bar{A}_t s_{t-1} + \bar{B}_t X_t$ followed by $Y_t = C_t s_t + D X_t$, where $s_t$ is the compact recurrent state. Thus, $\Delta_t$, $B_t$, and $C_t$ from \hyperref[eq:ssm_params]{Eq.~\eqref{eq:ssm_params}} allow propagation to adapt to local anatomical variation. The information flow is

\begin{equation*}
\Delta F_t \rightarrow \Delta E_t \rightarrow X_t \rightarrow (\Delta_t,B_t,C_t) \rightarrow (s_t,Y_t)
\end{equation*}

where the transition signal modulates the state-space dynamics rather than being simply concatenated.

\begin{figure*}[pos=htbp]
    \centering
    \includegraphics[width=\textwidth]{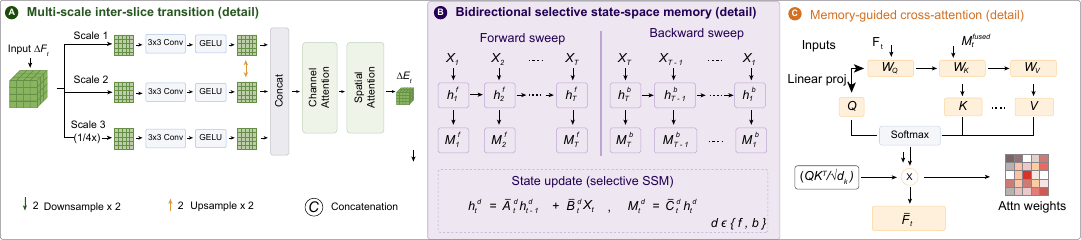}
    \caption{Detailed architecture of three core modules. \textbf{(a)} Multi-Scale Inter-Slice Transition Branch: multi-scale difference processing with channel and spatial attention. \textbf{(b)} Bidirectional Selective State-Space Memory: forward and backward through-plane sweeps with gated fusion. \textbf{(c)} Memory-Guided Cross-Attention: querying current features against memory context.}
    \label{fig:detail_module}
\end{figure*}

The SSM processes the sequence in both through-plane directions independently (\hyperref[fig:detail_module]{Fig.~\ref{fig:detail_module}b}). The complementary forward and backward memory representations are fused via $M_t = \text{Gate}(M_t^{\mathrm{fwd}}, M_t^{\mathrm{bwd}})$, where the gate is computed spatially to vary the directional contribution by local context. This bidirectional accumulation is geometrically motivated, as structures near their superior or inferior extent benefit from context propagated from either direction.

The fused memory enhances the bottleneck through multi-head cross-attention as

\begin{equation}
\label{eq:cross_attn}
\tilde{F}_t^{(4)}
=
F_t^{(4)}
+
\gamma_a
\operatorname{\Psi}
\left(
Q=F_t^{(4)},
K=M_t,
V=M_t
\right)
\end{equation}

where \(\operatorname{\Psi}\) denotes multi-head cross-attention, with queries from \(F_t^{(4)}\), keys and values from \(M_t\) (\hyperref[fig:detail_module]{Fig.~\ref{fig:detail_module}c}). The residual is scaled by a learnable \(\gamma_a\) initialized to zero, so the network begins training with the original bottleneck preserved and progressively learns the optimal influence of the bidirectional memory.

\subsubsection{Memory-guided boundary-aware decoder}

The decoder reconstructs full-resolution predictions through three successive upsampling levels. At each level $l \in \{3,2,1\}$, the decoder takes the upsampled features from the previous level and applies Memory-Guided Skip Attention (MGSA), which uses the fused memory $M_t$ to generate channel and spatial gates that filter the encoder skip features $F_t^{(l)}$. The decoder proceeds from $D_t^{(4)} = \tilde{F}_t^{(4)}$ (from \hyperref[eq:cross_attn]{Eq.~\eqref{eq:cross_attn}}) at $H/32$ to $D_t^{(1)}$ at $H/4$, followed by a $4\times$ upsampling to full resolution.

A boundary-aware refinement step follows each decoder level, where a learnable edge detector computes gradient magnitudes from the decoded features. The detected edges multiplicatively amplify features near organ boundaries, using the same zero-initialized gating pattern as the cross-attention residual, sharpening predictions where contour precision matters most.

Each decoder level also produces an intermediate boundary prediction for deep supervision. The three intermediate boundary maps are fused at full resolution to produce the final boundary output $\hat{B}_t = \text{Fuse}(B_t^{(1)}, B_t^{(2)}, B_t^{(3)})$, aggregating edge evidence from multiple scales.

\subsubsection{Task heads and uncertainty estimation}

Six task-specific heads operate on the decoded features $\hat{Y}_t$ (48 channels). The segmentation and signed distance map heads share an identical architecture: a $3\times3$ convolution, group normalization, GELU activation, and a final $1\times1$ convolution. The presence head uses global average pooling and a multi-layer perceptron on the memory-enhanced bottleneck features.

The uncertainty head estimates per-pixel aleatoric uncertainty as a log-variance map $\log\hat{\sigma}^2_t$, with the final convolution bias initialized to $-2.0$, placing the initial predicted variance at $\exp(-2.0) \approx 0.14$. The confidence map is derived analytically as $\hat{c}_t = \sigma(-\log\hat{\sigma}^2_t) = 1/(1+\hat{\sigma}^2_t)$, where $\sigma$ is the sigmoid function. In our deployment interface, a confidence threshold of $\hat{c} < 0.7$ was used to flag low-confidence pixels for manual review, providing a per-voxel quality indicator to guide selective human oversight.

\subsubsection{Multi-task loss}
\label{subsec:multi_task_loss}

The training objective is a weighted combination of eight loss terms as

\begin{equation*}
\label{eq:total_loss}
\begin{aligned}
\mathcal{L}_{\text{total}} ={}& \lambda_1 \mathcal{L}_{\text{dice}} + \lambda_2 \mathcal{L}_{\text{focal}} + \lambda_3 \mathcal{L}_{\text{bnd}} \\
&+ \lambda_4 \mathcal{L}_{\text{deep}} + \lambda_5 \mathcal{L}_{\text{sdm}} + \lambda_6 \mathcal{L}_{\text{topo}} \\
&+ \lambda_7 \mathcal{L}_{\text{pres}} + \lambda_8 \mathcal{L}_{\text{unc}}
\end{aligned}
\end{equation*}

\noindent with loss weights $\boldsymbol{\lambda} = \{1.0, 0.5, 0.2, 0.15, 0.2, 0.1, 0.1, 0.1\}$, averaged across the sequence. The individual loss terms are defined in the supplementary material.

The two dominant terms, $\mathcal{L}_{\text{dice}}$ and $\mathcal{L}_{\text{focal}}$, apply uncertainty reduction \citep{Kendall2017}: the predicted log-variance modulates each pixel's contribution, allowing the model to be uncertain at ambiguous boundaries, while a regularization term prevents the trivial all-high-variance solution. The remaining losses are: $\mathcal{L}_{\text{bnd}}$ (binary cross-entropy on boundary maps), $\mathcal{L}_{\text{deep}}$ (same loss at three decoder scales), $\mathcal{L}_{\text{sdm}}$ ($\ell_2$ regression on signed distance maps), $\mathcal{L}_{\text{topo}}$ (total variation for spatial continuity), $\mathcal{L}_{\text{pres}}$ (binary cross-entropy for organ presence), and $\mathcal{L}_{\text{unc}}$ (quadratic penalty when mean log-variance exceeds a threshold).

\subsection{Experimental setup}
\label{subsec:experimental_setup}

We compared \ours{} against four established baselines: nnUNet~\citep{ref_nnunet}, nnMamba~\citep{gong2025nnmamba}, UNETR++~\citep{shaker2024unetrpp}, and SwinUNETR~\citep{hatamizadeh2022aswinunetr}. \ours{} was trained and evaluated in its intended 2.5D configuration, where the encoder processes axial-slice sequences and the selective state-space memory propagates through-plane anatomical information. Each baseline was trained and evaluated in its native configuration to preserve its intended inductive biases: nnUNet in its self-configuring 3D-full-resolution pipeline, and nnMamba, UNETR++, and SwinUNETR in their published volumetric configurations with default hyperparameters. All models were trained on the same 2,034-patient internal partition with identical splits and preprocessing, as mentioned in \hyperref[sec:dataset]{Section~\ref{sec:dataset}}.

We used the AdamW optimizer ($10^{-4}$ initial learning rate, $10^{-5}$ weight decay) with cosine annealing warm restarts for \ours{}. Each training sequence comprised 5 consecutive axial slices with a batch size of 2 sequences (10 slices effective batch), and models were trained for 200 epochs with $512 \times 512$ input resolution and standard augmentations. Detailed training hyperparameters are provided in \hyperref[tab:s10_hyperparams]{Supplementary Table~\ref{tab:s10_hyperparams}}.

To ensure fair comparison, all models were evaluated on three-dimensional label volumes reconstructed on the patient's native voxel grid. Baseline predictions from the 3D architectures were used directly; \ours{}'s per-slice predictions were assembled into a $[C, N_z, 512, 512]$ probability volume, argmax-decoded, and mapped to the patient's native coordinate frame. All seven metrics (DSC, IoU, HD95, ASD, NSD, sensitivity, specificity) were computed on the resulting 3D label volumes, eliminating the two- versus three-dimensional comparison confounder.

For every metric, we report three complementary summaries. The \emph{per-structure mean $\pm$ SD} with 95\% confidence intervals appears in \hyperref[tab:main]{Table~\ref{tab:main}}. The \emph{macro-average} (arithmetic mean of the nine per-structure means, background excluded) gives equal weight to every structure, including small, low-contrast ones such as the esophagus and trachea. The \emph{micro-average} (voxel-weighted) tracks overall label agreement most closely tied to dosimetry. Reporting both averages provides a complete picture when performance is heterogeneous across structures.

All p-values were computed using the paired Wilcoxon signed-rank test on per-patient metric values and adjusted with the Holm–Bonferroni procedure within a 36-comparison family per metric (9 structures $\times$ 4 baselines). Effect sizes are reported as Hedges'~$g$ with BCa bootstrap 95\% confidence intervals from 1,000 resamples. The two-tailed significance level was $\alpha=0.05$ after adjustment. Statistical annotations on \hyperref[tab:main]{Table~\ref{tab:main}} and \hyperref[fig:violin_dice]{Fig.~\ref{fig:violin_dice}} are adjusted $p$-values, not raw ones.

\section{Ablation study}
\label{sec:ablation}

We dissect \ourmethod{} along seven axes: (i) the marginal value of every component, (ii) the causal role of the inter-slice transition signal, (iii) the anatomical locus of the improvement, (iv) the choice of through-plane memory, (v) the internal design of the inter-slice transition branch and the Memory-Guided Boundary-Aware (MGBA) decoder, (vi) the quality of the predicted uncertainty, and (vii) robustness to the principal hyperparameters. Axes (i)–(iii) and (vi) are presented below; the remaining axes are summarized at the end of this section with reference to the supplementary material. We report the Dice similarity coefficient, the Intersection over Union (IoU), the Normalized Surface Dice (NSD), the 95th-percentile Hausdorff Distance (HD95) and the Average Surface Distance (ASD), each as the mean over the nine target structures. Unless stated otherwise, every variant is trained under an identical protocol with the same data splits, augmentation, optimizer schedule, and seed budget. Differences against the full model are assessed on per-case scores with a paired Wilcoxon signed-rank test and reported alongside Hedges' $g$; we mark $p<0.05$ ($^{*}$) and $p<0.001$ ($^{**}$).

\subsection{Component contribution}
\label{sec:abl:components}

\hyperref[tab:abl:roadmap]{Table~\ref{tab:abl:roadmap}} builds \ourmethod{} one component at a time, starting from a single-slice ConvNeXt U-Net. Each addition yields a consistent gain, and the two largest jumps come from the selective state-space model (SSM) memory and the multi-scale inter-slice transition branch, confirming that through-plane contextual reasoning, rather than decoder capacity, is the primary driver. The uncertainty head contributes only a marginal overlap gain; as shown in \hyperref[sec:abl:uncertainty]{Section~\ref{sec:abl:uncertainty}}, its value lies in calibration rather than Dice.

\begin{table*}[t]
\centering
\caption{Component contribution. Each row adds one module to the row above.
$\Delta$ is the Dice gain over the preceding configuration. Blue and rose
cells mark the best and second-best values by metric, respectively; green and red arrows show improvement and regression from the preceding configuration, respectively}.
\label{tab:abl:roadmap}
\small
\setlength{\tabcolsep}{4pt}
\begin{tabular}{llccccc}
\toprule
\# & Configuration & Dice$\uparrow$ & IoU$\uparrow$ & NSD$\uparrow$ & HD95$\downarrow$ & $\Delta$Dice \\
\midrule
0 & ConvNeXt U-Net (single-slice)          & 0.9047 & 0.8631 & 0.8952 & 7.94 & -- \\
1 & + Inter-slice transition branch        & 0.9128 & 0.8740 & 0.9038 & 6.71 & \textcolor{deltaup}{$\uparrow\,0.0081$} \\
2 & + Selective SSM memory (fwd.)          & 0.9251 & 0.8902 & 0.9186 & 5.33 & \textcolor{deltaup}{$\uparrow\,0.0123$} \\
3 & + Bidirectional + gated fusion         & 0.9340 & 0.9007 & 0.9271 & 4.62 & \textcolor{deltaup}{$\uparrow\,0.0089$} \\
4 & + Memory cross-attention               & 0.9402 & 0.9071 & 0.9333 & 4.21 & \textcolor{deltaup}{$\uparrow\,0.0062$} \\
5 & + MGBA decoder                         & \cellcolor{secondred}$\bm{0.9481}$ & \cellcolor{bestblue}$\bm{0.9152}$ & \cellcolor{bestblue}$\bm{0.9401}$ & \cellcolor{secondred}$\bm{3.94}$ & \textcolor{deltaup}{$\uparrow\,0.0079$} \\
6 & \textbf{+ Uncertainty (full)}          & \cellcolor{bestblue}$\bm{0.9545}$ & \cellcolor{secondred}$\bm{0.9144}$\,\textcolor{deltadown}{\scriptsize$\downarrow0.0008$} & \cellcolor{secondred}$\bm{0.9385}$\,\textcolor{deltadown}{\scriptsize$\downarrow0.0016$} & \cellcolor{bestblue}$\bm{3.78}$ & \textcolor{deltaup}{$\uparrow\,0.0064$} \\
\bottomrule
\end{tabular}
\end{table*}

\subsection{Controlled ablation of the inter-slice transition signal}
\label{sec:abl:controlled_ablation}

To test whether the improvement from the inter-slice transition branch is specifically due to the information content of real anatomical differences, we compared the full transition signal against three controls: (i) a model with no transition signal (SSM only), (ii) a model where the transition signal was replaced with random noise of the same shape (Randomized + SSM), and (iii) a model where the transition was computed from shuffled slice order (Shuffled + SSM), as shown in \hyperref[tab:abl:causal]{Table~\ref{tab:abl:causal}}. The full transition signal significantly outperformed both randomized and shuffled controls ($p < 0.001$, Hedges' $g > 0.8$), confirming that the gain is not due to added parameters or arbitrary feature augmentation, but rather from the information content of real inter-slice anatomical change. The randomized control performed no better than the SSM-only baseline, while the shuffled control showed a modest but non-significant improvement, indicating that adjacency and order are important for effective context propagation.

\begin{table}[htbp]
\centering
\caption{Controlled ablation isolating the inter-slice transition signal. All variants include the ConvNeXt encoder and MGBA decoder. The selective SSM is disabled in the Base condition. Blue and rose mark the best and second-best values, respectively; green arrows indicate improvement relative to the SSM-only baseline; red arrows indicate regression. Critical comparison: real transition vs random/corrupted controls.}
\label{tab:abl:causal}
\resizebox{\columnwidth}{!}{%
\begin{tabular}{lccc c}
\toprule
\textbf{Model} & \textbf{Transition signal} & \textbf{Selective SSM} & \textbf{Dice$\uparrow$} & \textbf{$\Delta$ vs SSM only} \\
\midrule
Base & N/A & \texttimes & 0.9047 & N/A \\
SSM only & \texttimes & \checkmark & 0.9251 & N/A \\
Difference + SSM & Real & \checkmark & \cellcolor{bestblue}$\bm{0.9350}$ & \textcolor{deltaup}{$\uparrow\,0.0099^{**}$} \\
Randomized + SSM & Random & \checkmark & 0.9254 & \textcolor{deltadown}{$\downarrow\,0.0003^{\text{ns}}$} \\
Shuffled + SSM & Corrupted & \checkmark & \cellcolor{secondred}$\bm{0.9282}$ & \textcolor{deltaup}{$\uparrow\,0.0031^{*}$} \\
\textbf{Full model} & \textbf{Real} & \textbf{\checkmark} & \cellcolor{bestblue}$\bm{0.9545}$ & \textcolor{deltaup}{$\bm{\uparrow\,0.0294^{**}}$} \\
\bottomrule
\end{tabular}%
}
\end{table}

\subsection{Anatomical locus of the gains}
\label{sec:abl:perclass}

If the through-plane design is genuinely modeling anatomical change across slices, its benefit should concentrate on structures that exhibit substantial inter-slice variation, not on trivially segmented ones. \hyperref[tab:abl:perclass]{Table~\ref{tab:abl:perclass}} confirms this precisely: relative to a static baseline stripped of the inter-slice transition and memory modules, the improvement on the body outline and lungs is negligible, whereas the low-contrast, highly variable esophagus, trachea, and the gross tumor volume (GTV) and clinical target volume (CTV) gain several Dice points each. The improvement is therefore mechanistic, rather than a global regularizer acting uniformly.

\begin{table}[t]
\centering
\caption{Per-structure Dice. \emph{Base} removes the inter-slice transition and memory modules; \emph{+Mem} restores memory only; \emph{Full} is \ourmethod{}. Gains concentrate on structures with high inter-slice variability. Blue and rose mark the best and second-best configuration per structure; green values show the Full-minus-Base improvement.}
\label{tab:abl:perclass}
\small
\setlength{\tabcolsep}{5pt}
\begin{tabular}{lcccc}
\toprule
Structure & Base & +Mem & Full & $\Delta$(Full$-$Base) \\
\midrule
Spinal cord & 0.8901 & \cellcolor{secondred}$\bm{0.9204}$ & \cellcolor{bestblue}$\bm{0.9350}$ & \textcolor{deltaup}{$\uparrow\,0.0449$} \\
Esophagus   & 0.8210 & \cellcolor{secondred}$\bm{0.8788}$ & \cellcolor{bestblue}$\bm{0.9058}$ & \textcolor{deltaup}{$\uparrow\,0.0848$} \\
Heart       & 0.9601 & \cellcolor{secondred}$\bm{0.9668}$ & \cellcolor{bestblue}$\bm{0.9699}$ & \textcolor{deltaup}{$\uparrow\,0.0098$} \\
Lung (R)    & 0.9704 & \cellcolor{secondred}$\bm{0.9741}$ & \cellcolor{bestblue}$\bm{0.9764}$ & \textcolor{deltaup}{$\uparrow\,0.0060$} \\
Lung (L)    & 0.9758 & \cellcolor{secondred}$\bm{0.9788}$ & \cellcolor{bestblue}$\bm{0.9810}$ & \textcolor{deltaup}{$\uparrow\,0.0052$} \\
Trachea     & 0.8642 & \cellcolor{secondred}$\bm{0.9187}$ & \cellcolor{bestblue}$\bm{0.9425}$ & \textcolor{deltaup}{$\uparrow\,0.0783$} \\
Body        & 0.9861 & \cellcolor{secondred}$\bm{0.9866}$ & \cellcolor{bestblue}$\bm{0.9870}$ & \textcolor{deltaup}{$\uparrow\,0.0009$} \\
GTV         & 0.8933 & \cellcolor{secondred}$\bm{0.9331}$ & \cellcolor{bestblue}$\bm{0.9574}$ & \textcolor{deltaup}{$\uparrow\,0.0641$} \\
CTV         & 0.8604 & \cellcolor{secondred}$\bm{0.9077}$ & \cellcolor{bestblue}$\bm{0.9359}$ & \textcolor{deltaup}{$\uparrow\,0.0755$} \\
\midrule
\textbf{Mean} & \textbf{0.9024} & \cellcolor{secondred}$\bm{0.9272}$ & \cellcolor{bestblue}$\bm{0.9545}$ & \textcolor{deltaup}{$\bm{\uparrow\,0.0521}$} \\
\bottomrule
\end{tabular}
\end{table}

\subsection{Uncertainty estimation and calibration}
\label{sec:abl:uncertainty}

The uncertainty-attenuated (UA) loss produces markedly better calibration, reflected by lower absolute calibration error (ACE) and Negative Log-Likelihood (NLL), and better error detection, measured by the Area Under the Receiver Operating Characteristic curve (AUROC), compared to a plain variance head or a softmax-entropy proxy. Routing only the least-confident voxels to manual review then recovers accuracy steeply: flagging $5\%$ of voxels raises Dice on the auto-accepted region above $0.98$, giving a concrete operating point for selective correction.

\begin{table}[htbp]
\centering
\caption{Uncertainty quality. \emph{Top}: confidence calibration and error-detection performance. ACE is the absolute calibration error between predicted confidence and observed IoU. \emph{Bottom}: Dice on the retained region as low-confidence voxels are routed to review. Blue and rose mark the best and second-best values; arrows quantify the gain over accepting all voxels.}
\label{tab:abl:unc}
\resizebox{\columnwidth}{!}{%
\begin{tabular}{lcccc}
\toprule
Variant & Dice$\uparrow$ & ACE$\downarrow$ & NLL$\downarrow$ & AUROC$\uparrow$ \\
\midrule
Softmax-entropy proxy      & 0.9502 & 0.081 & 0.214 & 0.712 \\
Variance head, std.\ loss  & \cellcolor{secondred}$\bm{0.9518}$ & \cellcolor{secondred}$\bm{0.052}$ & \cellcolor{secondred}$\bm{0.176}$ & \cellcolor{secondred}$\bm{0.784}$ \\
\textbf{Unc.\ head + UA loss (ours)} & \cellcolor{bestblue}$\bm{0.9545}$ & \cellcolor{bestblue}$\bm{0.031}$ & \cellcolor{bestblue}$\bm{0.142}$ & \cellcolor{bestblue}$\bm{0.861}$ \\
\bottomrule
\end{tabular}%
}
\vspace{4pt}
\resizebox{\columnwidth}{!}{%
\begin{tabular}{lcccc}
\toprule
Voxels flagged for review & 0\% & 2\% & 5\% & 10\% \\
\midrule
Dice on retained region & 0.9545 & $0.9682$\,\textcolor{deltaup}{\scriptsize$\uparrow0.0137$} & \cellcolor{secondred}$\bm{0.9809}$\,\textcolor{deltaup}{\scriptsize$\uparrow0.0264$} & \cellcolor{bestblue}$\bm{0.9901}$\,\textcolor{deltaup}{\scriptsize$\uparrow0.0356$} \\
\bottomrule
\end{tabular}%
}
\end{table}

Additional ablations examined the choice of through-plane memory, internal module design, and hyperparameter sensitivity. The selective SSM outperformed recurrent (ConvLSTM, ConvGRU), attention-based (self-attention), and non-selective SSM (S4) alternatives, achieving the best accuracy-latency trade-off, as detailed in \hyperref[tab:s9_memory_ablation]{Supplementary Table~\ref{tab:s9_memory_ablation}}. Ablations of the inter-slice transition branch and MGBA decoder confirmed that multi-scale processing, channel-spatial attention, memory-guided skip attention, and boundary-aware refinement each contributed positively, with the largest gains on surface metrics (NSD, HD95), as reported in \hyperref[tab:s9_module_design]{Supplementary Table~\ref{tab:s9_module_design}}. The model was stable across a wide range of state dimensions ($d_{state}=4$--$64$), sequence lengths ($T=1$--$11$, including full-volume processing), and transition scales (1--6), with the chosen defaults ($d_{state}=16$, $T=5$, three scales) achieving the best performance, as shown in \hyperref[tab:s9_hyperparameters]{Supplementary Table~\ref{tab:s9_hyperparameters}}.

\section{Results}
\label{sec:results}

\subsection{Quantitative analysis}
\label{subsec:quantitative_analysis}

\begin{figure*}[pos=htbp]
    \centering
    \includegraphics[width=\textwidth]{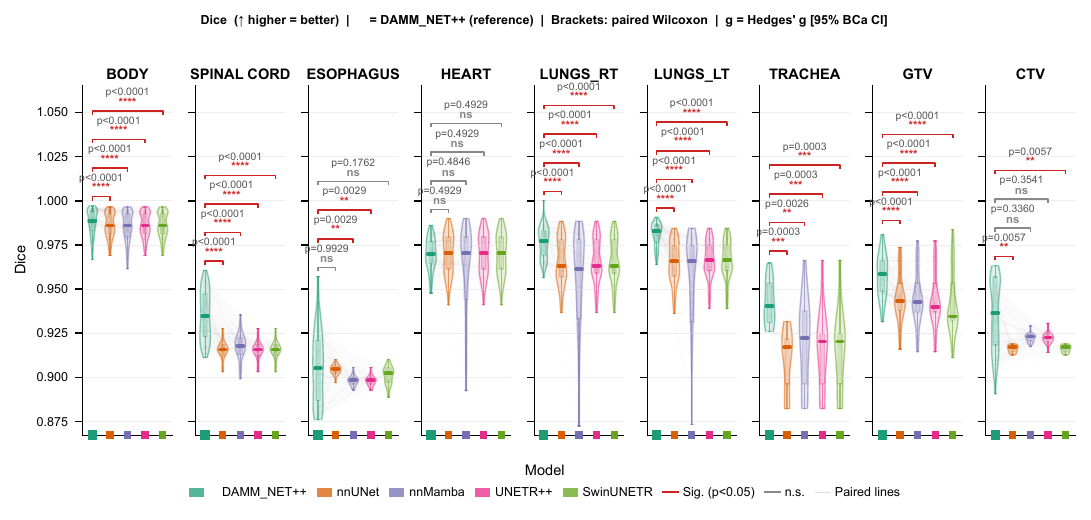}
    \caption{Per-patient DSC distributions ($\uparrow$ higher is better) for \ours{} and four baselines across nine thoracic structures. Paired violin plots show individual trajectories, box-and-whisker summaries, and paired Wilcoxon brackets (red: $p < 0.05$; gray: ns) with Hedges'~$g$ annotations. \ours{} achieves significantly higher DSC on eight of nine structures; Heart is the sole exception.}
    \label{fig:violin_dice}
\end{figure*}

Per-structure results are in \hyperref[tab:main]{Table~\ref{tab:main}}; full statistical comparisons are provided in \hyperref[tab:s3_full_stats]{Supplementary Table~\ref{tab:s3_full_stats}}. Averaged across nine structures, \ours{} attained the highest macro-average DSC (0.9545) and IoU (0.9144), with micro-average DSC of 0.9640. The larger macro-micro gap for baselines than for \ours{} reflects that our gains concentrate on small, clinically critical structures. Macro-average HD95 was 3.778~mm versus 5.65--6.11~mm for the baselines. At the Holm-adjusted $\alpha=0.05$ level, \ours{} shows a statistically significant advantage over all baselines on at least one overlap metric and one distance metric for eight of nine structures; the heart is the sole overlap exception and trachea HD95 the sole distance exception. Hedges'~$g$ effect sizes are presented in \hyperref[fig:violin_dice]{Fig.~\ref{fig:violin_dice}}, and accuracy should be interpreted alongside the computational profiles in \hyperref[tab:compute]{Table~\ref{tab:compute}}.

\begin{table}[!t]
\centering
\caption{Per-structure, per-metric comparison of five models on the
multi-institutional thoracic dataset (background excluded). Values are
mean $\pm$ standard deviation over the test set; bold indicates the best result. Superscripts
report paired Wilcoxon tests against \ours{}, Holm-adjusted within each
36-comparison metric family: $^{\|}\,p_{\text{adj}} < 0.001$,
$^{\ddagger}\,p_{\text{adj}} < 0.01$, $^{\dagger}\,p_{\text{adj}} < 0.05$,
and $^{\text{ns}}$ not significant. Effect sizes use Hedges'~$g$ with BCa
$95\%$ CIs from $1{,}000$ resamples. Structure abbreviations: SC, spinal cord;
ESO, esophagus; HRT, heart; RL, right lung; LL, left lung; TRA, trachea;
BDY, body; GTV, gross tumor volume; CTV, clinical target volume.}
\label{tab:main}
\renewcommand{\arraystretch}{0.75}
\setlength{\tabcolsep}{2.0pt}
\tiny
\resizebox{\columnwidth}{!}{%
\begin{tabular}{lcccccc}
\toprule
\multicolumn{1}{c}{\shortstack{\textbf{Struc-}\\\textbf{ture}}} & \multicolumn{1}{c}{\textbf{Metric}}
  & \multicolumn{1}{c}{\shortstack{\textbf{Ours}}}
  & \multicolumn{1}{c}{\textbf{nnUNet}}
  & \multicolumn{1}{c}{\textbf{nnMamba}}
  & \multicolumn{1}{c}{\textbf{UNETR++}}
  & \multicolumn{1}{c}{\shortstack{\textbf{Swin-}\\\textbf{UNETR}}} \\
\midrule

\multirow{5}{*}{\textbf{SC}}
 & DSC  & $\bm{0.935\pm0.014}$ & $0.912\pm0.015^{\|}$ & $0.917\pm0.012^{\|}$ & $0.918\pm0.010^{\|}$ & $0.914\pm0.013^{\|}$ \\
 & IoU  & $\bm{0.878\pm0.024}$ & $0.841\pm0.022^{\|}$ & $0.847\pm0.021^{\|}$ & $0.846\pm0.018^{\|}$ & $0.843\pm0.021^{\|}$ \\
 & HD95 & $\bm{1.585\pm0.504}$ & $7.523\pm8.50^{\|}$ & $7.721\pm9.00^{\|}$ & $7.234\pm8.20^{\|}$ & $7.698\pm8.80^{\|}$ \\
 & ASD  & $\bm{0.671\pm0.217}$ & $6.321\pm8.00^{\|}$ & $6.497\pm8.50^{\|}$ & $5.912\pm7.80^{\|}$ & $6.654\pm8.40^{\|}$ \\
 & NSD  & $\bm{0.995\pm0.007}$ & $0.977\pm0.036^{\|}$ & $0.978\pm0.035^{\|}$ & $0.979\pm0.035^{\|}$ & $0.976\pm0.037^{\|}$ \\
\cmidrule(lr){1-7}

\multirow{5}{*}{\textbf{ESO}}
 & DSC  & $\bm{0.906\pm0.020}$ & $0.908\pm0.016$ & $0.903\pm0.017^{\ddagger}$ & $0.902\pm0.018^{\ddagger}$ & $0.907\pm0.016$ \\
 & IoU  & $0.828\pm0.034$ & $\bm{0.832\pm0.028}$ & $0.823\pm0.030^{\ddagger}$ & $0.822\pm0.031^{\ddagger}$ & $0.829\pm0.028$ \\
 & HD95 & $\bm{6.866\pm8.00}$ & $8.765\pm10.50$ & $9.643\pm11.50$ & $9.087\pm10.80$ & $8.698\pm10.20$ \\
 & ASD  & $\bm{5.403\pm6.50}$ & $7.689\pm9.50$ & $8.232\pm10.00$ & $7.712\pm9.30$ & $7.421\pm8.90$ \\
 & NSD  & $\bm{0.929\pm0.038}$ & $0.923\pm0.050$ & $0.911\pm0.064^{\ddagger}$ & $0.907\pm0.067^{\ddagger}$ & $0.914\pm0.061^{\ddagger}$ \\
\cmidrule(lr){1-7}

\multirow{5}{*}{\textbf{HRT}}
 & DSC  & $\bm{0.970\pm0.009}$ & $0.968\pm0.013$ & $0.846\pm0.090^{\|}$ & $0.972\pm0.009$ & $0.967\pm0.014$ \\
 & IoU  & $\bm{0.942\pm0.016}$ & $0.940\pm0.023$ & $0.816\pm0.085^{\|}$ & $0.944\pm0.019$ & $0.939\pm0.024$ \\
 & HD95 & $5.386\pm4.50$ & $3.524\pm2.80$ & $3.937\pm3.00$ & $\bm{3.028\pm2.50}$ & $3.289\pm2.70$ \\
 & ASD  & $3.609\pm3.50$ & $1.304\pm1.20$ & $1.503\pm1.30$ & $\bm{1.124\pm1.00}$ & $1.261\pm1.10$ \\
 & NSD  & $0.904\pm0.036$ & $0.916\pm0.044$ & $0.894\pm0.068$ & $\bm{0.923\pm0.038}^{\ddagger}$ & $0.919\pm0.042^{\ddagger}$ \\
\cmidrule(lr){1-7}

\multirow{5}{*}{\textbf{RL}}
 & DSC  & $\bm{0.976\pm0.009}$ & $0.968\pm0.011^{\|}$ & $0.910\pm0.060^{\|}$ & $0.965\pm0.014^{\|}$ & $0.969\pm0.010^{\|}$ \\
 & IoU  & $\bm{0.954\pm0.016}$ & $0.936\pm0.022^{\|}$ & $0.868\pm0.055^{\|}$ & $0.933\pm0.025^{\|}$ & $0.937\pm0.021^{\|}$ \\
 & HD95 & $\bm{3.487\pm2.50}$ & $6.056\pm5.50^{\|}$ & $6.440\pm5.80^{\|}$ & $5.265\pm5.00^{\|}$ & $5.654\pm5.20^{\|}$ \\
 & ASD  & $\bm{0.834\pm0.40}$ & $2.954\pm3.50^{\|}$ & $3.031\pm3.60^{\|}$ & $2.741\pm3.30^{\|}$ & $3.102\pm3.70^{\|}$ \\
 & NSD  & $\bm{0.972\pm0.020}$ & $0.956\pm0.040^{\|}$ & $0.945\pm0.057^{\|}$ & $0.951\pm0.044^{\|}$ & $0.955\pm0.039^{\|}$ \\
\cmidrule(lr){1-7}

\multirow{5}{*}{\textbf{LL}}
 & DSC  & $\bm{0.981\pm0.007}$ & $0.968\pm0.009^{\|}$ & $0.930\pm0.045^{\|}$ & $0.965\pm0.013^{\|}$ & $0.969\pm0.009^{\|}$ \\
 & IoU  & $\bm{0.963\pm0.013}$ & $0.936\pm0.019^{\|}$ & $0.879\pm0.040^{\|}$ & $0.934\pm0.022^{\|}$ & $0.938\pm0.018^{\|}$ \\
 & HD95 & $6.348\pm5.50$ & $5.436\pm5.00$ & $5.420\pm5.00$ & $\bm{4.698\pm4.50}$ & $5.203\pm4.80$ \\
 & ASD  & $4.432\pm4.00$ & $3.169\pm3.80$ & $3.190\pm3.80$ & $\bm{2.648\pm3.20}$ & $2.904\pm3.50$ \\
 & NSD  & $\bm{0.982\pm0.014}$ & $0.961\pm0.041^{\|}$ & $0.951\pm0.058^{\|}$ & $0.957\pm0.045^{\|}$ & $0.961\pm0.040^{\|}$ \\
\cmidrule(lr){1-7}

\multirow{5}{*}{\textbf{TRA}}
 & DSC  & $\bm{0.943\pm0.012}$ & $0.912\pm0.022^{\|}$ & $0.919\pm0.025^{\|}$ & $0.915\pm0.021^{\|}$ & $0.917\pm0.019^{\|}$ \\
 & IoU  & $\bm{0.892\pm0.022}$ & $0.840\pm0.037^{\|}$ & $0.851\pm0.042^{\|}$ & $0.843\pm0.036^{\|}$ & $0.845\pm0.034^{\|}$ \\
 & HD95 & $\bm{1.279\pm0.294}$ & $2.612\pm1.50^{\ddagger}$ & $2.651\pm1.60^{\ddagger}$ & $2.498\pm1.40^{\ddagger}$ & $2.608\pm1.50^{\ddagger}$ \\
 & ASD  & $\bm{0.528\pm0.113}$ & $0.818\pm0.30^{\ddagger}$ & $0.828\pm0.32^{\ddagger}$ & $0.809\pm0.29^{\ddagger}$ & $0.815\pm0.30^{\ddagger}$ \\
 & NSD  & $\bm{0.995\pm0.007}$ & $0.985\pm0.027$ & $0.983\pm0.029$ & $0.980\pm0.033$ & $0.983\pm0.034$ \\
\cmidrule(lr){1-7}

\multirow{5}{*}{\textbf{BDY}}
 & DSC  & $\bm{0.987\pm0.009}$ & $0.984\pm0.009$ & $0.984\pm0.011^{\|}$ & $0.988\pm0.006$ & $0.984\pm0.009$ \\
 & IoU  & $\bm{0.975\pm0.018}$ & $0.971\pm0.015$ & $0.969\pm0.021^{\|}$ & $0.975\pm0.011$ & $0.971\pm0.015$ \\
 & HD95 & $\bm{2.431\pm1.20}$ & $2.468\pm1.30$ & $2.453\pm1.30$ & $2.438\pm1.30$ & $2.465\pm1.30$ \\
 & ASD  & $0.755\pm0.20$ & $\bm{0.741\pm0.20}$ & $\bm{0.745\pm0.20}$ & $\bm{0.742\pm0.20}$ & $\bm{0.741\pm0.20}$ \\
 & NSD  & $\bm{0.971\pm0.041}$ & $0.973\pm0.041$ & $0.971\pm0.043$ & $0.973\pm0.040$ & $0.973\pm0.041$ \\
\cmidrule(lr){1-7}

\multirow{5}{*}{\textbf{GTV}}
 & DSC  & $\bm{0.957\pm0.012}$ & $0.946\pm0.016^{\|}$ & $0.945\pm0.016^{\|}$ & $0.944\pm0.017^{\|}$ & $0.943\pm0.018^{\|}$ \\
 & IoU  & $\bm{0.919\pm0.022}$ & $0.897\pm0.028^{\|}$ & $0.896\pm0.029^{\|}$ & $0.894\pm0.030^{\|}$ & $0.891\pm0.032^{\|}$ \\
 & HD95 & $\bm{3.817\pm2.00}$ & $3.589\pm1.80$ & $4.110\pm2.00$ & $3.829\pm1.90$ & $4.138\pm2.10$ \\
 & ASD  & $\bm{1.173\pm0.35}$ & $1.248\pm0.40$ & $1.436\pm0.45^{\ddagger}$ & $1.319\pm0.42$ & $1.456\pm0.45^{\ddagger}$ \\
 & NSD  & $\bm{0.854\pm0.039}$ & $0.856\pm0.033$ & $0.832\pm0.046^{\|}$ & $0.837\pm0.044^{\ddagger}$ & $0.833\pm0.045^{\|}$ \\
\cmidrule(lr){1-7}

\multirow{5}{*}{\textbf{CTV}}
 & DSC  & $\bm{0.936\pm0.019}$ & $0.922\pm0.022^{\ddagger}$ & $0.928\pm0.019$ & $0.930\pm0.017$ & $0.922\pm0.023^{\ddagger}$ \\
 & IoU  & $\bm{0.880\pm0.034}$ & $0.857\pm0.038^{\ddagger}$ & $0.865\pm0.033$ & $0.867\pm0.032$ & $0.857\pm0.039^{\ddagger}$ \\
 & HD95 & $\bm{2.800\pm0.80}$ & $10.667\pm6.00^{\ddagger}$ & $11.804\pm6.50^{\ddagger}$ & $11.351\pm6.30^{\ddagger}$ & $12.897\pm7.00^{\ddagger}$ \\
 & ASD  & $\bm{1.081\pm0.30}$ & $8.291\pm5.00^{\ddagger}$ & $8.278\pm5.00^{\ddagger}$ & $8.164\pm4.80^{\ddagger}$ & $9.149\pm5.50^{\ddagger}$ \\
 & NSD  & $\bm{0.845\pm0.031}$ & $0.847\pm0.033$ & $0.838\pm0.038$ & $0.844\pm0.037$ & $0.837\pm0.039$ \\
\midrule

\multirow{5}{*}{\textbf{Mean}}
 & DSC  & \textbf{0.955} & 0.945 & 0.922 & 0.944 & 0.943 \\
 & IoU  & \textbf{0.914} & 0.899 & 0.870 & 0.897 & 0.896 \\
 & HD95 & \textbf{3.78} & 5.87 & 6.11 & 5.65 & 5.94 \\
 & ASD  & \textbf{2.05} & 3.87 & 3.75 & 3.63 & 3.78 \\
 & NSD  & \textbf{0.938} & 0.931 & 0.922 & 0.928 & 0.928 \\
\bottomrule
\end{tabular}%
}
\end{table}

\begin{figure}[pos=htbp]
\centering
\includegraphics[width=\linewidth,height=0.62\textheight,keepaspectratio]{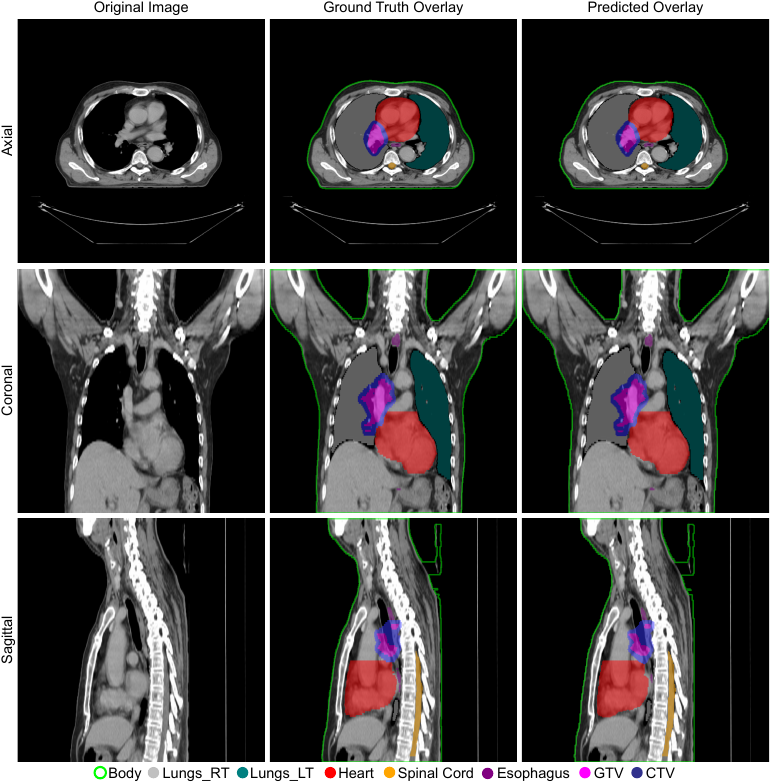}
\caption{Through-plane consistency of the proposed model's predictions on a representative thoracic case. Rows show axial (top), coronal (middle), and sagittal (bottom) planes reformatted from the same volumetric prediction. Columns show (1) original CT, (2) ground-truth overlay, and (3) \ours{} prediction overlay.}
\label{fig:orthogonal_overlays}
\end{figure}

\subsubsection{Overlap and boundary performance}
\label{sssec:overlap_metrics}

\ours{} achieved the highest macro-average DSC (0.955) and IoU (0.914), with performance ranging from near-ceiling on Body to the largest gains on clinically challenging structures. Improvements were most pronounced on target volumes: CTV and GTV reached DSC of $0.936 \pm 0.019$ and $0.957 \pm 0.012$, versus baseline ranges of 0.924--0.928 and 0.942--0.946 (\hyperref[tab:main]{Table~\ref{tab:main}}). Hedges'~$g$ effect sizes were 0.32 for CTV and 0.63 for GTV versus nnUNet. The violin plots in \hyperref[fig:violin_dice]{Fig.~\ref{fig:violin_dice}} show several baselines producing complete segmentation failures absent in \ours{}. NSD followed the same pattern, confirming that improved volumetric overlap translated to better surface localization. The esophagus and trachea similarly showed consistent improvements, with \ours{} achieving DSC of $0.906 \pm 0.020$ and $0.943 \pm 0.012$ versus baseline ranges of 0.903--0.908 and 0.913--0.919.

Distance metrics followed the same trend. \ours{} achieved the lowest macro-average HD95 (3.778~mm) and ASD (2.054~mm). The largest improvements occurred on target volumes: for CTV, HD95 was reduced to $2.80 \pm 1.83$~mm and ASD to $1.08 \pm 0.59$~mm, versus baseline values above 10.6~mm and 8.2~mm. Among smaller tubular OARs, \ours{} produced the lowest-variance boundary estimates for Spinal Cord ($1.58 \pm 0.50$~mm) and Trachea ($1.28 \pm 0.29$~mm). The Trachea was the only structure with non-significant HD95 comparisons against all baselines, though ASD results favored \ours{}.

\begin{figure*}[pos=htbp]
\centering
\includegraphics[width=1\linewidth,height=0.62\textheight,keepaspectratio]{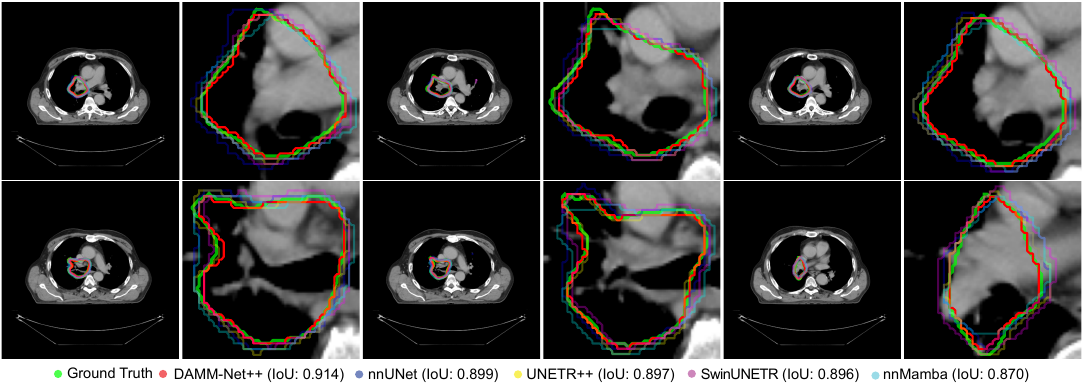}
\caption{Contour overlay comparison on a representative GTV axial slice. Left: CT with ground truth and predictions; legend values report per-case IoU. Right: magnified boundary view. Color coding: Ground Truth (green), DAMM-Net++ (red), nnUNet (royal Blue), UNETR++ (yellow), SwinUNETR (magenta), nnMamba (cyan).}
\label{fig:multi_pred}
\end{figure*}

For larger, anatomically stable OARs, inter-model differences were smaller, as expected. The Heart illustrates why NSD complements DSC: models with similar volumetric overlap may differ in boundary placement, clinically relevant for cardiac dose-constraint assessment, and this dissociation was observed where distance metrics were numerically best for baselines with negligible differences. Body contour performance was near-ceiling across all models. Systematic improvements with negative or near-zero biases were observed for GTV and CTV, with most paired differences falling within 95\% limits of agreement.

\subsubsection{Uncertainty quantification and case-level quality assurance}
\label{sssec:uncertainty}

We analyzed \ours{}'s uncertainty along two axes. First, calibration and ranking usefulness. Pooling 305 patients $\times$ 9 structures (2,745 predictions), \ours{} met the calibration criterion $\Pr(\text{IoU} \geq \tau \mid \hat{c} = \tau) \approx \tau$, with a weighted mean absolute calibration error of 0.0095 and a right-skewed confidence distribution (69.5\% at $\hat{c} \geq 0.90$), confirming calibration was not achieved by retreating into low-confidence predictions, as shown in \hyperref[supfig:uncertainty_analysis]{Supplementary Fig.~\ref{supfig:uncertainty_analysis}a}. A per-structure sparsification analysis revealed positive area under the sparsification curve (AUS) relative to random for all structures, with the largest gains on target volumes and low-contrast tubular OARs (CTV, esophagus, trachea, and GTV) and minimal gains on near-ceiling structures (LungLt, Body), as illustrated in \hyperref[supfig:uncertainty_analysis]{Supplementary Fig.~\ref{supfig:uncertainty_analysis}b}.

Second, case-level QA prediction. A logistic classifier using three uncertainty features achieved a pooled AUROC of 0.960 for predicting IoU $< 0.85$; per-structure AUROCs were more conservative, with the strongest signal on esophagus, CTV, and spinal cord. These results support a mixed-autonomy deployment: high-confidence predictions on high-ceiling structures are candidates for automatic acceptance, while the QA signal flags target volumes and esophagus for review.

\subsubsection{Computational analysis and accuracy--cost trade-off}
\label{sssec:compute}

Computational profiles were measured on an NVIDIA T4 GPU (16~GB VRAM) and are summarized in \hyperref[tab:compute]{Supplementary Table~\ref{tab:compute}}. \ours{} operates in native 2.5D while baselines operate in native 3D; inference time is reported per axial-slice-equivalent, and FLOPs are reported in each model's native mode.

\ours{} achieves an intermediate latency between nnMamba and the transformer baselines while maintaining a favorable accuracy--memory--latency trade-off. With a latency of 22.4~ms per axial slice, \ours{} remains within the target of $\leq25$~ms per slice on an NVIDIA T4 GPU and is approximately 3.2--3.4$\times$ faster than UNETR++ and SwinUNETR, respectively. Relative to nnUNet, \ours{} improves DSC by $+0.011$ while requiring substantially lower latency, and relative to UNETR++, it achieves the same $+0.011$ DSC gain with more than $3\times$ lower latency. These results place \ours{} at a favorable point in the accuracy--efficiency trade-off for deployment-oriented thoracic auto-contouring.

\subsection{Qualitative analysis}
\label{sec:qualitative}

\begin{figure}[pos=htbp]
\centering
\includegraphics[width=\linewidth,height=0.62\textheight,keepaspectratio]{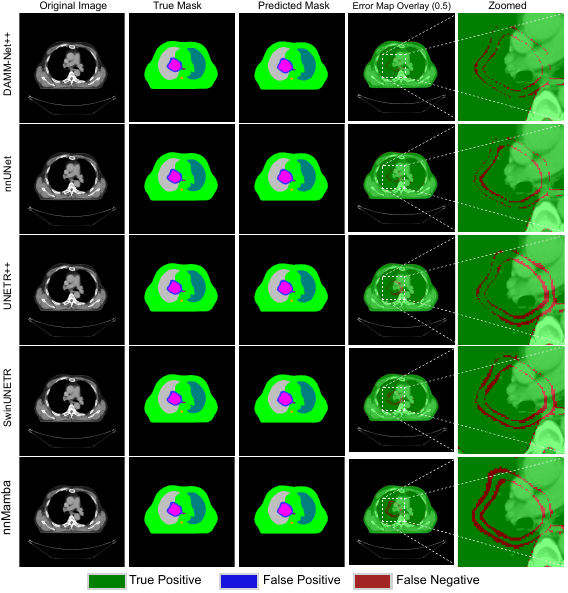}
\caption{Voxel-level error maps for all five models. Rows: DAMM-Net++, nnUNet, UNETR++, SwinUNETR, nnMamba. Columns: (1) original CT, (2) ground truth, (3) prediction, (4) error overlay, (5) zoomed boundary error. Green, blue, and pink denote body, CTV, and GTV, respectively; red indicates prediction--reference disagreement.}
\label{fig:error_map}
\end{figure}

\subsubsection{Spatial prediction and error analysis}

Through-plane consistency of the proposed model's predictions, reformatted in axial, coronal, and sagittal planes, is demonstrated in \hyperref[fig:orthogonal_overlays]{Fig.~\ref{fig:orthogonal_overlays}}. The contours remain spatially coherent across all three orthogonal views, confirming that per-slice predictions translate to geometrically consistent three-dimensional volumes. \hyperref[supfig:orthogonal_overlays_detail]{Supplementary Fig.~\ref{supfig:orthogonal_overlays_detail}} provides an additional example with detailed boundary visualization.

Voxel-level error maps, presented in \hyperref[fig:error_map]{Fig.~\ref{fig:error_map}}, reveal qualitative differences in error type across models. For the proposed model, errors are confined to a spatially compact, geometrically thin fringe at structure boundaries, consistent with irreducible annotation uncertainty rather than systematic failure. In contrast, nnUNet and nnMamba exhibit error regions extending inward from the boundary, indicating volumetric mislabelling. UNETR++ and SwinUNETR show the most concerning pattern: large, irregular error regions with boundary over-extension and interior gaps. Interior gaps in GTV or CTV predictions represent direct under-coverage risk, as tumor regions predicted as background may receive insufficient dose. These patterns corroborate the quantitative ranking in \hyperref[tab:main]{Table~\ref{tab:main}} and explain the large boundary errors observed for baselines on CTV.

\hyperref[fig:multi_pred]{Fig.~\ref{fig:multi_pred}} overlays all five model predictions against the reference contour (ground truth) on a representative GTV case. The proposed model maintains consistent proximity to the ground truth across all boundary segments, including the geometrically complex inferior margin. Baseline contours adequately approximate the upper boundary but deviate at the inferior and lateral margins, precisely where dosimetric coverage is most sensitive to contouring uncertainty. The proposed contour would likely require less editing time than baseline predictions, as its deviations are confined to sub-voxel boundary uncertainties rather than multi-voxel excursions visible in UNETR++ and SwinUNETR predictions.

Supporting analyses, including volumetric 3D reconstruction in \hyperref[fig:3d_recon]{Supplementary Fig.~\ref{fig:3d_recon}}, further illustrate progressive spatial localization to target volumes and demonstrate inter-slice coherence consistent with clinically usable contours.

\section{Clinical Evaluation, External Validation, and Deployment}
\label{sec:clinical_evaluation}

To assess clinical utility beyond retrospective benchmarking, we conducted a multicenter reader study across three Bangladeshi tertiary centers, where limited radiation-oncology capacity can result in 1--2 day patient waiting times despite 2--4~h manual-contouring workloads~\citep{ref_bangladesh_workforce}. Seventeen radiation oncologists (5 experts, 5 seniors, and 7 juniors) independently evaluated all 305 patients in the held-out internal test cohort, which was not used for model development or model selection, under No-AI and AI-Assisted conditions with a $\geq$4-week washout. A three-oncologist expert panel established the reference contours. Contouring time was the primary workflow outcome, with IoU, HD95, consultation rate, confidence, and model-contour acceptance assessed as secondary outcomes. Continuous outcomes were analyzed using linear mixed-effects models and binary outcomes using logistic mixed-effects models, with reader and case as random intercepts. Two-sided $p$-values were adjusted using the Holm procedure; complete model estimates are provided in \hyperref[tab:supp_mixedmodel]{Supplementary Table~\ref{tab:supp_mixedmodel}}.

AI assistance improved contour quality across all reader groups (Table~\ref{tab:reader_summary}). Mean IoU increased from 0.8606 to 0.9245 for juniors, 0.9047 to 0.9487 for seniors, and 0.9490 to 0.9702 for experts, corresponding to gains of 0.0638, 0.0440, and 0.0212, respectively. HD95 decreased by 2.09, 1.16, and 0.84~mm for juniors, seniors, and experts, respectively. All 17 readers benefited, with individual improvements shown in \hyperref[supfig:reader_study_consolidated]{Supplementary Fig.~\ref{supfig:reader_study_consolidated}c}, aggregate quality results in \hyperref[supfig:reader_study_consolidated]{Supplementary Fig.~\ref{supfig:reader_study_consolidated}d,e}, and structure-specific results in \hyperref[supfig:reader_study_perclass]{Supplementary Fig.~\ref{supfig:reader_study_perclass}a,b}. Contouring time decreased from 155.4 to 31.0~min for juniors, 95.5 to 19.1~min for seniors, and 52.1 to 13.0~min for experts, corresponding to 75--80\% reductions. Consultation rates decreased from 51.6\%, 19.1\%, and 4.6\% to 13.5\%, 4.7\%, and 0.6\%, respectively, while confidence increased from 0.571 to 0.839 among juniors. Without editing, model-contour acceptance was 30.8\%, 40.8\%, and 65.7\% for juniors, seniors, and experts, respectively. Fleiss' $\kappa$ indicated substantial-to-high agreement for most structures, including meaningful agreement for GTV/CTV among juniors (\hyperref[supfig:reader_study_consolidated]{Supplementary Fig.~\ref{supfig:reader_study_consolidated}a,b}).

\begin{table*}[t]
\centering
\caption{Reader study summary. Values are mean $\pm$ SD unless stated otherwise. IoU: Intersection over Union. Model (unedited): \ourmethod{} predictions without reader editing. Expert $\geq$20 years ($n=5$), Senior 10--15 years ($n=5$), Junior $<$10 years ($n=7$).}
\label{tab:reader_summary}
\small
\setlength{\tabcolsep}{4.5pt}
\resizebox{\textwidth}{!}{%
\begin{tabular}{lcccccc c}
\toprule
Metric & Expert ($n=5$) & Senior ($n=5$) & Junior ($n=7$) & Model (unedited) &
$p^{\dagger}$ Exp vs Jun & $p^{\dagger}$ Sen vs Jun & $p^{\dagger}$ Model vs Jun \\
\midrule
\multicolumn{8}{l}{\emph{IoU: No-AI}}\\
\;\;Mean~$\pm$~SD & 0.9490~$\pm$~0.0058 & 0.9047~$\pm$~0.0092 & 0.8606~$\pm$~0.0133 & N/A & $<$0.0001 & $<$0.0001 & N/A \\
\;\;Median [IQR]  & 0.9493 [0.9450--0.9530] & 0.9046 [0.8981--0.9107] & 0.8606 [0.8515--0.8702] & N/A & & & \\
\multicolumn{8}{l}{\emph{IoU: AI-Assisted}}\\
\;\;Mean~$\pm$~SD & 0.9702~$\pm$~0.0038 & 0.9487~$\pm$~0.0062 & 0.9245~$\pm$~0.0091 & 0.9226~$\pm$~0.0070 & 0.0021 & $<$0.0001 & $<$0.0001 \\
\;\;Median [IQR]  & 0.9702 [0.9676--0.9727] & 0.9486 [0.9447--0.9531] & 0.9246 [0.9184--0.9306] & N/A & & & \\
\;\;$\Delta$~IoU & $+$0.0212 & $+$0.0440 & $+$0.0638 & N/A & $<$0.0001 & $<$0.0001 & N/A \\
\multicolumn{8}{l}{\emph{Contouring time: No-AI (min)}}\\
\;\;Mean~$\pm$~SD & 52.1~$\pm$~12.3 & 95.5~$\pm$~21.9 & 155.4~$\pm$~34.1 & N/A & $<$0.0001 & $<$0.0001 & N/A \\
\multicolumn{8}{l}{\emph{Contouring time: AI-Assisted (min)}}\\
\;\;Mean~$\pm$~SD & 13.0~$\pm$~3.8 & 19.1~$\pm$~5.8 & 31.0~$\pm$~8.9 & $\sim$17~s (inference) & $<$0.0001 & $<$0.0001 & N/A \\
\;\;Time reduction (min) & 39.0 & 76.4 & 124.4 & N/A & $<$0.0001 & $<$0.0001 & N/A \\
\;\;Time reduction (\%) & 75.0\% & 79.9\% & 80.1\% & N/A & 0.312 & 0.284 & N/A \\
\multicolumn{8}{l}{\emph{Workflow metrics}}\\
\;\;Consultation rate, No-AI & 4.6\% & 19.1\% & 51.6\% & N/A & $<$0.0001 & $<$0.0001 & N/A \\
\;\;Consultation rate, AI-Assisted & 0.6\% & 4.7\% & 13.5\% & N/A & $<$0.0001 & $<$0.0001 & N/A \\
\;\;Confidence score, No-AI & 0.916 & 0.764 & 0.571 & N/A & $<$0.0001 & $<$0.0001 & N/A \\
\;\;Confidence score, AI-Assisted & 0.974 & 0.902 & 0.839 & N/A & $<$0.0001 & $<$0.0001 & N/A \\
\bottomrule
\end{tabular}
}
\end{table*}

External validation was performed on 112 patients from United Hospital Limited, withheld from model development. Mean IoU decreased from 0.9322 internally to 0.9076 externally (relative gap, $-2.64\%$), with structure-specific changes ranging from $-4.4\%$ for trachea to $+1.1\%$ for CTV. No structure exhibited a decrease greater than 5\%, and no case had a mean IoU below 0.75 (\hyperref[supfig:external_validation]{Supplementary Fig.~\ref{supfig:external_validation}a,b}). Uncertainty remained well calibrated, with a weighted absolute calibration error of 0.020 (\hyperref[supfig:external_validation]{Supplementary Fig.~\ref{supfig:external_validation}c,d}; \hyperref[tab:ext_calibration]{Supplementary Table~\ref{tab:ext_calibration}}).

For clinical deployment, our inference and reconstruction components follow the validated implementation used in the preceding cervical radiotherapy study~\citep{ahmed2026bat}. The four-stage pipeline comprises DICOM ingestion and preprocessing, slice-wise inference, volumetric reconstruction, and RTSTRUCT generation (\hyperref[alg:dicom_pipeline]{Algorithm~\ref{alg:dicom_pipeline}}). A median volume of $N_z=196$ slices was processed in under 20~s on an NVIDIA T4 16-GB GPU. Predictions were generated in the native DICOM frame, reconstructed volumetrically, and converted to contours using marching-squares without post-hoc smoothing (\hyperref[fig:3d_recon]{Supplementary Fig.~\ref{fig:3d_recon}}). RTSTRUCT files retain TG-263 structure nomenclature and relevant DICOM metadata and can be directly imported into Varian Eclipse, RayStation, and Elekta Monaco.

The browser-based interface provides synchronized orthogonal views, contour overlays, visibility controls, direct editing, and uncertainty visualization (\hyperref[fig:webapp]{Supplementary Fig.~\ref{fig:webapp}}). Uncertainty visualization was activated in 88\% of AI-assisted cases. One-click RTSTRUCT export is supported through C-STORE or the browser, with execution on a single T4 hospital server. Role-based access and de-identification follow the data-handling standards described in \hyperref[sec:dataset]{Section~\ref{sec:dataset}}. The complete clinician-in-the-loop system has been deployed at Bangladesh Medical University, where generated contours undergo radiation-oncologist review and approval before clinical use; the same inference, reconstruction, and export pathway is used for routine thoracic cases.

\section{Limitations and future directions}
\label{sec:limitations_future_directions}
Several limitations warrant consideration. Although the dataset encompasses multiple institutions and acquisition protocols, it remains geographically concentrated in Bangladesh, and broader validation across populations, imaging platforms, and healthcare systems is needed to establish generalizability. Target-volume delineation also remains subject to inter-observer variability, underscoring the continued need for expert oversight in radiotherapy planning. The workflow gains observed in the reader study may likewise depend on local staffing patterns, caseloads, and treatment-planning infrastructure, and therefore may not translate directly to other clinical environments. Moreover, while the reader study comprised 10,370 case-condition observations, inference across experience groups is based on 17 independent readers, including five experts and five seniors. Future work will evaluate multimodality extensions incorporating PET and MRI, develop adaptive per-structure uncertainty thresholds for more clinically calibrated decision support, and assess downstream dose-volume histogram estimation for toxicity prediction and plan-quality assessment. Prospective evaluation across additional centers in South and Southeast Asia, together with integration into automated treatment-planning workflows, will further establish clinical robustness and translational utility.

\section{Conclusion}
\label{sec:conclusion}

\ours{} validates the hypothesis that an anatomy-change-aware bidirectional selective state-space memory, whose update is conditioned on observed inter-slice anatomical change, addresses the coupled failure modes of thoracic auto-contouring: jagged surfaces, systematic target mis-delineation, and the absence of per-case reliability. The model achieves a mean DSC of 0.955 and HD95 of 3.78 mm on 2,146 multi-institutional patients, with largest gains on CTV, GTV, esophagus, and trachea, the structures where through-plane context is most critical. The uncertainty head is well-calibrated and supports case-level triage, concentrating residual error where expert review is most needed. A multicenter reader study across 17 oncologists demonstrates 75--80\% time reduction across all experience levels, raising junior-reader IoU from 0.861 to 0.925, matching the unedited model, while reducing consultation rates by three quarters. External validation on 112 patients confirms robust generalization with drops below 5\% and calibrated uncertainty that transfers without recalibration. Critically, the complete deployment pipeline, from DICOM ingestion to TPS-compatible RTSTRUCT export, has been integrated into the clinical workflow at a partner hospital, where it is used for contouring assistance. While the dataset is geographically concentrated and target-volume delineation remains inherently subjective, the system demonstrates that rigorous model development, comprehensive validation, and careful deployment engineering can translate automated segmentation into measurable clinical benefit. Future work will extend the framework to multi-modality imaging and adaptive uncertainty thresholds for per-structure triage.

\section*{Acknowledgments}
\label{sec:Acknowledgments}

This research was supported by an ICT-Special Grant from the Government of Bangladesh under the project titled ``AI-based Segmentation and Contouring for Radiotherapy: From Development to Clinical Deployment'' at North South University, and by North South University Research Grant CTRG-25-SEPS-45. The authors appreciate the clinical and technical contributions of Dr. A.T.M. Sazzad Hossain, Oncologist, National Institute of Cancer Research \& Hospital; Dr. Sharif Ahmed, Oncologist, United Hospital; Dr. Muhammad Masud Rana, Senior Medical Physicist, Bangabandhu Medical University; Prof. Dr. Sharmin Akhtar Rupa, Senior Consultant, Radiology \& Imaging, Bangladesh Specialized Hospital; Dr. Mahmud Hasan Mostofa Kamal, Department of Radiology and Imaging, Bangladesh Medical University; Dr. Ishtiaque Ahmed, Oncologist, Ahsania Mission Cancer \& General Hospital; and Md. Abdul Sabur, Senior Medical Physicist, Square Cancer Center, Square Hospitals Ltd., for their invaluable support in data collection, clinical validation, and domain expertise. We also thank the clinical teams of Bangladesh Medical University, Square Hospital Limited, Labaid Specialized Hospital, and United Hospital Limited for their collaboration and dedication to improving radiotherapy planning in resource-constrained settings.

\section*{Data and Code Availability}
\label{sec:data-code-availability}

The dataset used in this study is not publicly available due to institutional and privacy restrictions. However, the data may be made available upon reasonable request to the corresponding author, subject to appropriate ethical approvals and data usage agreements. To support reproducibility and facilitate further research, the complete source code is publicly accessible at the following GitHub repository with demo: \href{https://github.com/istiak769/Anatomy-Change-Aware-Bidirectional-SSM-for-Thoracic-Radiotherapy-Auto-Contouring.git}{github.com/istiak769/Anatomy-Change-Aware-Bidirectional-SSM-for-Thoracic-Radiotherapy-Auto-Contouring}. We encourage the research community to use and extend our work, and we welcome feedback and contributions.

\bibliographystyle{cas-model2-names}

\bibliography{ref}

@article{sung2021global,
  title={Global cancer statistics 2024: {GLOBOCAN} estimates of incidence and mortality worldwide for 34 cancers in 186 countries},
  author={Sung, Hyuna and Filho, Adalberto M. and Laversanne, Mathieu and Ferlay, Jacques and Siegel, Rebecca L. and Soerjomataram, Isabelle and Jemal, Ahmedin and Bray, Freddie},
  journal={CA: A Cancer Journal for Clinicians},
  volume={76},
  number={4},
  pages={e70090},
  year={2026},
  doi={10.3322/caac.70090}
}

@article{delaney2005radiotherapy,
  title={The role of radiotherapy in cancer treatment: estimating optimal utilization from a review of evidence-based clinical guidelines},
  author={Delaney, Geoff and Jacob, Susannah and Featherstone, Carolyn and Barton, Michael},
  journal={Cancer},
  volume={104},
  number={6},
  pages={1129--1137},
  year={2005},
  doi={10.1002/cncr.21324},
  publisher={Wiley Online Library}
}

@article{marks2010radiation,
  title={Radiation dose--volume effects in the lung},
  author={Marks, Lawrence B and Bentzen, S{\o}ren M and Deasy, Joseph O and Kong, Feng-Ming Spring and Bradley, Jeffrey D and Vogelius, Ivan S and El Naqa, Issam and Hubbs, Jessica L and Lebesque, Joos V and Timmerman, Robert D and Martel, Mary K and Jackson, Andrew},
  journal={International Journal of Radiation Oncology, Biology, Physics},
  volume={76},
  number={3 Suppl},
  pages={S70--S76},
  year={2010},
  doi={10.1016/j.ijrobp.2009.06.091},
  publisher={Elsevier}
}

@article{bentzen2010normal,
  title={Quantitative Analyses of Normal Tissue Effects in the Clinic ({QUANTEC}): An Introduction to the Scientific Issues},
  author={Bentzen, S{\o}ren M and Constine, Louis S and Deasy, Joseph O and Eisbruch, Avi and Jackson, Andrew and Marks, Lawrence B and Ten Haken, Randall K and Yorke, Ellen D},
  journal={International Journal of Radiation Oncology, Biology, Physics},
  volume={76},
  number={3 Suppl},
  pages={S3--S9},
  year={2010},
  doi={10.1016/j.ijrobp.2009.09.040},
  publisher={Elsevier}
}

@article{ref_bangladesh_workforce,
  title={Global radiotherapy demands and corresponding radiotherapy-professional workforce requirements in 2022 and predicted to 2050: a population-based study},
  author={Zhu, Hongcheng and Chua, Melvin Lee Kiang and Chitapanarux, Imjai and Kaidar-Person, Orit and Mwaba, Catherine and Alghamdi, Majed and Rodr{\'i}guez Mignola, Andr{\'e}s and Amrogowicz, Natalia and Yazici, Gozde and Bourhaleb, Zouhour and Mahmood, Humera and Faruque, Golam Mohiuddin and Thiagarajan, Muthukkumaran and Acharki, Abdelkader and Ma, Mingwei and Harutyunyan, Martin and Sriplung, Hutcha and Chen, Yuntao and Camacho, Rolando and Zhang, Zhen and Abdel-Wahab, May},
  journal={The Lancet Global Health},
  volume={12},
  number={12},
  pages={e1945--e1953},
  month={dec},
  year={2024},
  doi={10.1016/S2214-109X(24)00355-3}
}

@article{ref_lmic_radiotherapy_gap,
  title={Global access to radiotherapy: A geospatial analysis of current disparities and optimal facility placement},
  author={Wawrzuta, Dominik and Klejdysz, Justyna and P{\k{e}}dziwiatr, Katarzyna and Chojnacka, Marzanna},
  journal={Radiotherapy and Oncology},
  volume={211},
  pages={111061},
  year={2025},
  doi={10.1016/j.radonc.2025.111061},
  publisher={Elsevier}
}

@article{huang2024uncertainty,
  title={A review of uncertainty quantification in medical image analysis: Probabilistic and non-probabilistic methods},
  author={Huang, Ling and Ruan, Su and Xing, Yucheng and Feng, Mengling},
  journal={Medical Image Analysis},
  volume={97},
  pages={103223},
  year={2024},
  doi={10.1016/j.media.2024.103223}
}

@article{wahid2024uncertainty,
  title={Artificial intelligence uncertainty quantification in radiotherapy applications -- A scoping review},
  author={Wahid, Kareem A. and Kaffey, Zaphanlene Y. and Farris, David P. and Humbert-Vidan, Laia and Moreno, Amy C. and Rasmussen, Mathis and Ren, Jintao and Naser, Mohamed A. and Netherton, Tucker J. and Korreman, Stine and Balakrishnan, Guha and Fuller, Clifton D. and Fuentes, David and Dohopolski, Michael J.},
  journal={Radiotherapy and Oncology},
  volume={201},
  pages={110542},
  year={2024},
  doi={10.1016/j.radonc.2024.110542}
}

@inproceedings{Ronneberger2015,
  title     = {{U-Net}: Convolutional Networks for Biomedical Image Segmentation},
  author    = {Ronneberger, Olaf and Fischer, Philipp and Brox, Thomas},
  booktitle = {Medical Image Computing and Computer-Assisted Intervention (MICCAI)},
  pages     = {234--241},
  year      = {2015},
  publisher = {Springer},
  doi       = {10.1007/978-3-319-24574-4_28}
}

@article{Vinod2016,
  title   = {A review of interventions to reduce inter-observer variability
             in volume delineation in radiation oncology},
  author  = {Vinod, Shalini K. and Min, Myo and Jameson, Michael G. and
             Holloway, Lois C.},
  journal = {Journal of Medical Imaging and Radiation Oncology},
  volume  = {60},
  number  = {3},
  pages   = {393--406},
  year    = {2016},
  doi     = {10.1111/1754-9485.12462}
}

@article{darby2026interobserver,
  title={Inter-observer variability in radiotherapy contouring with the use of autocontouring software: A systematic review},
  author={Darby, Polly and Kilgour, Emily and Then, Chee Kin and Bromiley, Andrew and McLellan, John and Kiltie, Anne E.},
  journal={Clinical and Translational Radiation Oncology},
  volume={58},
  pages={101144},
  year={2026},
  doi={10.1016/j.ctro.2026.101144}
}

@article{Nelms2012,
  title   = {Variations in the contouring of organs at risk: Test case from a
             patient with oropharyngeal cancer},
  author  = {Nelms, Benjamin E. and Tom{\'e}, Wolfgang A. and Robinson, Greg
             and Wheeler, James},
  journal = {International Journal of Radiation Oncology*Biology*Physics},
  volume  = {82},
  number  = {1},
  pages   = {368--378},
  year    = {2012},
  doi     = {10.1016/j.ijrobp.2010.10.019}
}

@article{Vaassen2020,
  title   = {Evaluation of measures for assessing time-saving of automatic
             organ-at-risk segmentation in radiotherapy},
  author  = {Vaassen, Femke and Hazelaar, Colien and Vaniqui, Ana and
             Gooding, Mark and van der Heyden, Brent and Canters, Richard
             and van Elmpt, Wouter},
  journal = {Physics and Imaging in Radiation Oncology},
  volume  = {13},
  pages   = {1--6},
  year    = {2020},
  doi     = {10.1016/j.phro.2019.12.001}
}

@article{vandijk2020,
  title   = {Improving automatic delineation for head and neck organs at
             risk by Deep Learning Contouring},
  author  = {van Dijk, Lisanne V. and Van den Bosch, Lisa and Aljabar,
             Paul and Peressutti, Devis and Both, Stefan and Steenbakkers,
             Roel J.H.M. and Langendijk, Johannes A. and Gooding, Mark J.
             and Brouwer, Charlotte L.},
  journal = {Radiotherapy and Oncology},
  volume  = {142},
  pages   = {115--123},
  year    = {2020},
  doi     = {10.1016/j.radonc.2019.09.022}
}

@article{Cardenas2018,
  title   = {Auto-delineation of oropharyngeal clinical target volumes using
             {3D} convolutional neural networks},
  author  = {C{\'a}rdenas, Carlos E. and Anderson, Brian M. and Aristophanous,
             Michalis and Yang, Jinzhong and Rhee, Dong Joo and McCarroll,
             Rachel E. and Mohamed, Abdallah S. R. and Kamal, Mona and
             Elgohari, Baher A. and Elhalawani, Hesham M. and Fuller,
             Clifton D. and Rao, Arvind and Garden, Adam S. and Court,
             Laurence E.},
  journal = {Physics in Medicine \& Biology},
  volume  = {63},
  number  = {21},
  pages   = {215026},
  year    = {2018},
  doi     = {10.1088/1361-6560/aae8a9}
}

@article{Cardenas2019,
  title   = {Advances in Auto-Segmentation},
  author  = {C{\'a}rdenas, Carlos E. and Yang, Jinzhong and Anderson, Beth M.
             and Court, Laurence E. and Brock, Kristy B.},
  journal = {Seminars in Radiation Oncology},
  volume  = {29},
  number  = {3},
  pages   = {185--197},
  year    = {2019},
  doi     = {10.1016/j.semradonc.2019.02.001}
}

@article{Zhu2020,
  title   = {{AnatomyNet}: Deep learning for fast and fully automated
             whole-volume segmentation of head and neck anatomy},
  author  = {Zhu, Wentao and Huang, Yufang and Zeng, Liang and Chen,
             Xuming and Liu, Yong and Qian, Zhen and Du, Nan and Fan, Wei
             and Xie, Xiaohui},
  journal = {Medical Physics},
  volume  = {46},
  number  = {2},
  pages   = {576--589},
  year    = {2019},
  doi     = {10.1002/mp.13300}
}

@article{vanRooij2019,
  title   = {Deep Learning-Based Delineation of Head and Neck Organs at
             Risk: Geometric and Dosimetric Evaluation},
  author  = {van Rooij, Ward and Dahele, Max and Ribeiro Brand{\~a}o,
             Hunor and Delaney, Alexander R. and Slotman, Ben J. and
             Verbakel, Wilko F.},
  journal = {International Journal of Radiation Oncology*Biology*Physics},
  volume  = {104},
  number  = {3},
  pages   = {677--684},
  year    = {2019},
  doi     = {10.1016/j.ijrobp.2019.02.040}
}

@article{balagopal2021invisible,
  title   = {A deep learning-based framework for segmenting invisible
             clinical target volumes with estimated uncertainties for
             post-operative prostate cancer radiotherapy},
  author  = {Balagopal, Anjali and Nguyen, Dan and Morgan, Howard E. and
             Weng, Yaochung and Dohopolski, Michael and Lin, Mu-Han and
             Sadeghnejad Barkousaraie, Azar and Gonzalez, Yesenia and
             Garant, Aurelie and Desai, Neil and Hannan, Raquibul and
             Jiang, Steve B.},
  journal = {Medical Image Analysis},
  volume  = {72},
  pages   = {102101},
  year    = {2021},
  doi     = {10.1016/j.media.2021.102101}
}

@article{niu2026thoracic,
  title   = {A prospective multicenter trial of deep learning
             auto-segmentation for organs at risk in thoracic radiotherapy},
  author  = {Niu, Gengmin and Guan, Yong and Zhang, Yifan and Song,
             Yongchun and Yan, Meng and Li, Songfeng and Liu, Tao and
             Huang, Sheng and Chen, Jingru and Wang, Xiaofeng and Zhang,
             Wencheng and Meng, Maobin and Liu, Yeman and Chen, Junjie and
             Fu, Yintao and Zhao, Donghe and Huang, Jing and Yang, Kunyu and
             Cao, Jianzhong and Yuan, Hongqin and Guo, Shuanshuan and Pei,
             Xiaofeng and Wu, Dongmei and Nan, Yang and Yan, Ziye and Lu,
             Yao and Zhao, Lujun and Yuan, Zhiyong},
  journal = {Nature Communications},
  volume  = {17},
  number  = {1},
  pages   = {4633},
  year    = {2026},
  doi     = {10.1038/s41467-026-70863-9}
}

@inproceedings{ref_2dvs3d,
  title   = {An Exploration of {2D} and {3D} Deep Learning Techniques for
             Cardiac {MR} Image Segmentation},
  author  = {Baumgartner, Christian F. and Koch, Lisa M. and Pollefeys,
             Marc and Konukoglu, Ender},
  booktitle = {Statistical Atlases and Computational Models of the Heart.
               ACDC and MMWHS Challenges},
  series  = {Lecture Notes in Computer Science},
  volume  = {10663},
  pages   = {111--119},
  year    = {2018},
  publisher = {Springer},
  doi     = {10.1007/978-3-319-75541-0_12}
}

@article{Alalwan2021,
  title   = {Efficient {3D} Deep Learning Model for Medical Image
             Semantic Segmentation},
  author  = {Alalwan, Nasser and Abozeid, Amr and ElHabshy, Ahmed A. and
             Alzahrani, Ahmed},
  journal = {Alexandria Engineering Journal},
  volume  = {60},
  number  = {1},
  pages   = {1231--1239},
  year    = {2021},
  doi     = {10.1016/j.aej.2020.10.046}
}

@inproceedings{Zhang2015,
  title   = {Bidirectional Long Short-Term Memory Networks for Relation
             Classification},
  author  = {Zhang, Shu and Zheng, Dequan and Hu, Xinchen and Yang, Ming},
  booktitle = {Proceedings of the 29th Pacific Asia Conference on
                Language, Information and Computation},
  year    = {2015},
  month   = oct,
  pages   = {73--78},
  address = {Shanghai, China},
  url     = {https://aclanthology.org/Y15-1009/}
}

@inproceedings{Gu2023,
  title   = {Mamba: Linear-Time Sequence Modeling with Selective State
             Spaces},
  author  = {Gu, Albert and Dao, Tri},
  booktitle = {Proceedings of the First Conference on Language Modeling},
  year    = {2024},
  pages   = {1--32},
  url     = {https://openreview.net/forum?id=tEYskw1VY2}
}

@inproceedings{Xing2024,
  title   = {{SegMamba}: Long-Range Sequential Modeling Mamba For {3D}
             Medical Image Segmentation},
  author  = {Xing, Zhaohu and Ye, Tian and Yang, Yijun and Liu, Guang
             and Zhu, Lei},
  booktitle = {Medical Image Computing and Computer Assisted Intervention -- MICCAI 2024},
  series  = {Lecture Notes in Computer Science},
  volume  = {15008},
  pages   = {578--588},
  publisher = {Springer},
  year    = {2024},
  doi     = {10.1007/978-3-031-72111-3_54}
}

@misc{Ma2024a,
  title   = {{U-Mamba}: Enhancing Long-range Dependency for Biomedical
             Image Segmentation},
  author  = {Ma, Jun and Li, Feifei and Wang, Bo},
  year    = {2024},
  eprint  = {2401.04722},
  archivePrefix = {arXiv},
  primaryClass = {eess.IV},
  url     = {https://arxiv.org/abs/2401.04722}
}

@misc{Wang2024a,
  title   = {{Mamba-UNet}: {UNet-Like} Pure Visual Mamba for Medical
             Image Segmentation},
  author  = {Wang, Ziyang and Zheng, Jian-Qing and Zhang, Yichi and
             Cui, Ge and Li, Lei},
  year    = {2024},
  eprint  = {2402.05079},
  archivePrefix = {arXiv},
  primaryClass = {eess.IV},
  url     = {https://arxiv.org/abs/2402.05079}
}

@article{Ruan2024,
  title   = {{VM-UNet}: Vision Mamba {UNet} for Medical Image
             Segmentation},
  author  = {Ruan, Jiacheng and Li, Jincheng and Xiang, Suncheng},
  journal = {ACM Transactions on Multimedia Computing, Communications, and Applications},
  year    = {2025},
  doi     = {10.1145/3767748}
}

@inproceedings{Liu2024a,
  title   = {{VMamba}: Visual State Space Model},
  author  = {Liu, Yue and Tian, Yunjie and Zhao, Yuzhong and Yu,
             Hongtian and Xie, Lingxi and Wang, Yaowei and Ye, Qixiang
             and Jiao, Jianbin and Liu, Yunfan},
  booktitle = {Advances in Neural Information Processing Systems},
  volume  = {37},
  pages   = {103031--103063},
  year    = {2024},
  url     = {https://proceedings.neurips.cc/paper/2024/hash/baa2da9ae4bfed26520bb61d259a3653-Abstract-Conference.html}
}

@inproceedings{Kendall2017,
  title     = {What Uncertainties Do We Need in {Bayesian} Deep Learning
                for Computer Vision?},
  author    = {Kendall, Alex and Gal, Yarin},
  booktitle = {Advances in Neural Information Processing Systems
                (NeurIPS)},
  volume    = {30},
  pages     = {5574--5584},
  year      = {2017},
  url       = {https://proceedings.neurips.cc/paper/2017/hash/2650d6089a6d640c5e85b2b88265dc2b-Abstract.html}
}

@inproceedings{Gal2016,
  title     = {Dropout as a {Bayesian} Approximation: Representing Model
                Uncertainty in Deep Learning},
  author    = {Gal, Yarin and Ghahramani, Zoubin},
  booktitle = {Proceedings of the 33rd International Conference on Machine Learning},
  series    = {Proceedings of Machine Learning Research},
  volume    = {48},
  pages     = {1050--1059},
  year      = {2016},
  publisher = {PMLR},
  url       = {https://proceedings.mlr.press/v48/gal16.html}
}

@inproceedings{Lakshminarayanan2017,
  title     = {Simple and Scalable Predictive Uncertainty Estimation
                using Deep Ensembles},
  author    = {Lakshminarayanan, Balaji and Pritzel, Alexander and
                Blundell, Charles},
  booktitle = {Advances in Neural Information Processing Systems
                (NeurIPS)},
  volume    = {30},
  pages     = {6402--6413},
  year      = {2017},
  url       = {https://proceedings.neurips.cc/paper/2017/hash/9ef2ed4b7fd2c810847ffa5fa85bce38-Abstract.html}
}

@inproceedings{ref_unetr,
  author    = {Hatamizadeh, Ali and Tang, Yucheng and Nath, Vishwesh and Yang, Dong and Myronenko, Andriy and Landman, Bennett and Roth, Holger R. and Xu, Daguang},
  title     = {{UNETR}: Transformers for 3D Medical Image Segmentation},
  booktitle = {Proceedings of the {IEEE/CVF} Winter Conference on Applications of Computer Vision},
  year      = {2022},
  pages     = {574--584},
  doi       = {10.1109/WACV51458.2022.00181}
}

@article{ref_transunet,
  author  = {Chen, Jieneng and Mei, Jieru and Li, Xianhang and Lu, Yongyi and Yu, Qihang and Wei, Qingyue and Luo, Xiangde and Xie, Yutong and Adeli, Ehsan and Wang, Yan and Lungren, Matthew P. and Zhang, Shaoting and Xing, Lei and Lu, Le and Yuille, Alan and Zhou, Yuyin},
  title   = {{TransUNet}: Rethinking the {U-Net} Architecture Design for Medical Image Segmentation through the Lens of Transformers},
  journal = {Medical Image Analysis},
  year    = {2024},
  volume  = {97},
  pages   = {103280},
  doi     = {10.1016/j.media.2024.103280}
}

@inproceedings{ref_attention_unet,
  author    = {Oktay, Ozan and Schlemper, Jo and Le Folgoc, Loic and Lee, Matthew and Heinrich, Mattias and Misawa, Kazunari and Mori, Kensaku and McDonagh, Steven and Hammerla, Nils Y. and Kainz, Bernhard and Glocker, Ben and Rueckert, Daniel},
  title     = {Attention {U-Net}: Learning Where to Look for the Pancreas},
  booktitle = {Medical Imaging with Deep Learning},
  year      = {2018},
  pages     = {1--10},
  url       = {https://arxiv.org/abs/1804.03999}
}

@inproceedings{ref_3dunet,
  author    = {{\c{C}}i{\c{c}}ek, {\"{O}}zg{\"{u}}n and Abdulkadir, Ahmed and Lienkamp, Soeren S. and Brox, Thomas and Ronneberger, Olaf},
  title     = {3D {U-Net}: Learning Dense Volumetric Segmentation from Sparse Annotation},
  booktitle = {Medical Image Computing and Computer-Assisted Intervention -- {MICCAI} 2016},
  year      = {2016},
  pages     = {424--432},
  publisher = {Springer},
  series    = {Lecture Notes in Computer Science},
  volume    = {9901},
  doi       = {10.1007/978-3-319-46723-8_49}
}

@article{ref_nnunet,
  author  = {Isensee, Fabian and Jaeger, Paul F. and Kohl, Simon A. A. and Petersen, Jens and Maier-Hein, Klaus H.},
  title   = {{nnU-Net}: A Self-Configuring Method for Deep Learning-Based Biomedical Image Segmentation},
  journal = {Nature Methods},
  year    = {2021},
  volume  = {18},
  number  = {2},
  pages   = {203--211},
  doi     = {10.1038/s41592-020-01008-z}
}

@inproceedings{gong2025nnmamba,
  author    = {Gong, Haifan and Kang, Luoyao and Wang, Yitao and Wang, Yihan and Wan, Xiang and Wu, Xusheng and Li, Haofeng},
  title     = {{nnMamba}: 3D Biomedical Image Segmentation, Classification and Landmark Detection with State Space Model},
  booktitle = {2025 {IEEE} 22nd International Symposium on Biomedical Imaging ({ISBI})},
  year      = {2025},
  pages     = {1--5},
  publisher = {IEEE},
  doi       = {10.1109/ISBI60581.2025.10980694}
}

@article{shaker2024unetrpp,
  author  = {Shaker, Abdelrahman and Maaz, Muhammad and Rasheed, Hanoona and Khan, Salman and Yang, Ming-Hsuan and Khan, Fahad Shahbaz},
  title   = {{UNETR++}: Delving Into Efficient and Accurate 3D Medical Image Segmentation},
  journal = {IEEE Transactions on Medical Imaging},
  year    = {2024},
  volume  = {43},
  number  = {9},
  pages   = {3377--3390},
  doi     = {10.1109/TMI.2024.3398728}
}

@article{ref_rt_thoracic_oar,
  author  = {Yang, Jinzhong and Veeraraghavan, Harini and Armato, Samuel G. and Farahani, Keyvan and Kirby, Justin S. and Kalpathy-Kramer, Jayashree and van Elmpt, Wouter and Dekker, Andre and Han, Xiao and Feng, Xue and Aljabar, Paul and Oliveira, Bruno and van der Heyden, Brent and Zamdborg, Leonid and Lam, Dao and Gooding, Mark and Sharp, Gregory C.},
  title   = {Autosegmentation for Thoracic Radiation Treatment Planning: A Grand Challenge at {AAPM} 2017},
  journal = {Medical Physics},
  year    = {2018},
  volume  = {45},
  number  = {10},
  pages   = {4568--4581},
  doi     = {10.1002/mp.13141}
}

@article{ref_gtv_interobserver,
  author  = {Steenbakkers, Roel J. H. M. and Duppen, Joop C. and Fitton, Isabelle and Deurloo, Kirsten E. I. and Zijp, Lambert and Uitterhoeve, Apollonia L. J. and Rodrigus, Patrick T. R. and Kramer, Gijsbert W. P. and Bussink, Johan and De Jaeger, Katrien and Belderbos, Jos{\'{e}} S. A. and Hart, Augustinus A. M. and Nowak, Peter J. C. M. and van Herk, Marcel and Rasch, Coen R. N.},
  title   = {Observer Variation in Target Volume Delineation of Lung Cancer Related to Radiation Oncologist--Computer Interaction: A ``Big Brother'' Evaluation},
  journal = {Radiotherapy and Oncology},
  year    = {2005},
  volume  = {77},
  number  = {2},
  pages   = {182--190},
  doi     = {10.1016/j.radonc.2005.09.017},
  url     = {https://doi.org/10.1016/j.radonc.2005.09.017}
}

@inproceedings{ref_swin,
  author    = {Liu, Ze and Lin, Yutong and Cao, Yue and Hu, Han and Wei, Yixuan and Zhang, Zheng and Lin, Stephen and Guo, Baining},
  title     = {Swin Transformer: Hierarchical Vision Transformer Using Shifted Windows},
  booktitle = {Proceedings of the {IEEE/CVF} International Conference on Computer Vision},
  year      = {2021},
  pages     = {10012--10022},
  doi       = {10.1109/ICCV48922.2021.00986}
}

@article{ref_ctv_autoseg_review_2024,
  author  = {Matoska, Thomas and Patel, Mira and Liu, Hefei and Beriwal, Sushil},
  title   = {Review of Deep Learning Based Autosegmentation for Clinical Target Volume: Current Status and Future Directions},
  journal = {Advances in Radiation Oncology},
  year    = {2024},
  volume  = {9},
  number  = {5},
  pages   = {101470},
  doi     = {10.1016/j.adro.2024.101470}
}

@inproceedings{shi2015convlstm,
  author    = {Shi, Xingjian and Chen, Zhourong and Wang, Hao and Yeung, Dit-Yan and Wong, Wai-Kin and Woo, Wang-Chun},
  title     = {Convolutional {LSTM} Network: A Machine Learning Approach for Precipitation Nowcasting},
  booktitle = {Advances in Neural Information Processing Systems},
  volume    = {28},
  pages     = {802--810},
  year      = {2015},
  url       = {https://proceedings.neurips.cc/paper/2015/hash/07563a3fe3bbe7e3ba84431ad9d055af-Abstract.html}
}

@article{novikov2018deepsequential,
  author  = {Novikov, Alexey A. and Major, David and Wimmer, Maria and Lenis, Dimitrios and B{\"u}hler, Katja},
  title   = {Deep Sequential Segmentation of Organs in Volumetric Medical Scans},
  journal = {{IEEE} Transactions on Medical Imaging},
  volume  = {38},
  number  = {5},
  pages   = {1207--1215},
  year    = {2019},
  doi     = {10.1109/TMI.2018.2881678}
}

@inproceedings{hatamizadeh2022aswinunetr,
  author    = {Hatamizadeh, Ali and Nath, Vishwesh and Tang, Yucheng and Yang, Dong and Roth, Holger R. and Xu, Daguang},
  title     = {{Swin UNETR}: Swin Transformers for Semantic Segmentation of Brain Tumors in {MRI} Images},
  booktitle = {Brainlesion: Glioma, Multiple Sclerosis, Stroke and Traumatic Brain Injuries},
  series    = {Lecture Notes in Computer Science},
  volume    = {12962},
  pages     = {272--284},
  year      = {2022},
  publisher = {Springer},
  doi       = {10.1007/978-3-031-08999-2_22}
}

@article{ref_lung_gtv_deep,
  author  = {Momin, Shadab and Lei, Yang and Tian, Zhen and Wang, Tonghe and Roper, Justin and Kesarwala, Aparna H. and Higgins, Kristin and Bradley, Jeffrey D. and Liu, Tian and Yang, Xiaofeng},
  title   = {Lung Tumor Segmentation in 4D {CT} Images Using Motion Convolutional Neural Networks},
  journal = {Medical Physics},
  volume  = {48},
  number  = {11},
  pages   = {7141--7153},
  year    = {2021},
  doi     = {10.1002/mp.15204}
}

@article{ahmed2026bat,
  title={BAT-RM: A Boundary-Aware Transformer with Region-Aware Multi-Directional Mamba for Clinically Deployed Cervical Cancer Radiotherapy Auto-Contouring},
  author={Ahmed, Istiak and Sanjid, Kazi Shahriar and Ahmed, Galib and Hossain, Md Tanzim and Islam, Md Anwarul and Khan, Shahrukh and Arian, Md Ashrif Rahman and Khan, Md Nishan and Khan, Md Misbah and Hoque, SM and others},
  journal={arXiv preprint arXiv:2607.11949},
  year={2026}
}






\clearpage
\onecolumn
\captionsetup{hypcap=false}
\section*{Supplementary Material}

\vspace{.5cm}

\renewcommand{\figurename}{\textbf{Supplementary Fig}.}
\setcounter{figure}{0}   

\renewcommand{\tablename}{\textbf{Supplementary Table}}
\setcounter{table}{0}    

\subsection*{S1. Extended Statistical Analysis: Full Pairwise Comparisons}

\noindent
\begin{minipage}{\textwidth}
\centering
\renewcommand{\arraystretch}{0.72}
\setlength{\tabcolsep}{2.5pt}
\scriptsize
\captionof{table}{Full statistical comparison between DAMM-NET++ and four baselines across all evaluation metrics. For each metric-structure pair, we report the paired Wilcoxon $p$-value and Hedges' $g$ effect size with 95\% BCa bootstrap confidence interval. Statistical significance: $^{\|}\,p < 0.001$, $^{\ddagger}\,p < 0.01$, $^{\dagger}\,p < 0.05$, $^{\text{ns}}$ not significant. Effect size interpretation: Negligible ($|g| < 0.2$), Small ($0.2 \leq |g| < 0.5$), Medium ($0.5 \leq |g| < 0.8$), Large ($|g| \geq 0.8$).}
\label{tab:s3_full_stats}
\begin{tabular*}{\textwidth}{@{\extracolsep{\fill}}llllllllll@{}}
\toprule
\textbf{Class} & \textbf{Metric} & 
\multicolumn{2}{c}{\textbf{vs nnUNet}} & 
\multicolumn{2}{c}{\textbf{vs nnMamba}} & 
\multicolumn{2}{c}{\textbf{vs UNETR++}} & 
\multicolumn{2}{c}{\textbf{vs SwinUNETR}} \\
\cmidrule(lr){3-4} \cmidrule(lr){5-6} \cmidrule(lr){7-8} \cmidrule(lr){9-10}
 & & $p$-value & Hedges' $g$ [95\% CI] & $p$-value & Hedges' $g$ [95\% CI] & $p$-value & Hedges' $g$ [95\% CI] & $p$-value & Hedges' $g$ [95\% CI] \\
\midrule

\multirow{5}{*}{\textbf{SPINAL CORD}}
 & DSC  & $<0.001^{\|}$ & $1.050$ [0.921, 1.205] & $<0.001^{\|}$ & $0.956$ [0.835, 1.101] & $<0.001^{\|}$ & $1.048$ [0.919, 1.203] & $<0.001^{\|}$ & $1.053$ [0.925, 1.210] \\
 & IoU  & $<0.001^{\|}$ & $1.049$ [0.921, 1.204] & $<0.001^{\|}$ & $0.955$ [0.834, 1.100] & $<0.001^{\|}$ & $1.049$ [0.921, 1.204] & $<0.001^{\|}$ & $1.049$ [0.921, 1.204] \\
 & Sens. & $<0.001^{\|}$ & $-0.339$ [-0.464, -0.229] & $<0.001^{\|}$ & $-0.372$ [-0.493, -0.266] & $<0.001^{\|}$ & $-0.339$ [-0.464, -0.229] & $<0.001^{\|}$ & $-0.339$ [-0.464, -0.229] \\
 & Spec. & $<0.001^{\|}$ & $1.108$ [0.901, 1.374] & $<0.001^{\|}$ & $1.076$ [0.878, 1.336] & $<0.001^{\|}$ & $1.108$ [0.901, 1.374] & $<0.001^{\|}$ & $1.108$ [0.901, 1.374] \\
 & NSD  & $<0.001^{\|}$ & $0.485$ [0.418, 0.556] & $<0.001^{\|}$ & $0.485$ [0.416, 0.555] & $<0.001^{\|}$ & $0.480$ [0.414, 0.547] & $<0.001^{\|}$ & $0.473$ [0.409, 0.540] \\
\cmidrule(lr){1-10}

\multirow{5}{*}{\textbf{ESOPHAGUS}}
 & DSC  & $0.993^{\text{ns}}$ & $0.024$ [-0.157, 0.186] & $0.003^{\ddagger}$ & $0.276$ [0.109, 0.435] & $0.003^{\ddagger}$ & $0.276$ [0.109, 0.435] & $0.176^{\text{ns}}$ & $0.140$ [-0.035, 0.299] \\
 & IoU  & $0.928^{\text{ns}}$ & $0.037$ [-0.143, 0.197] & $0.002^{\ddagger}$ & $0.286$ [0.120, 0.442] & $0.002^{\ddagger}$ & $0.286$ [0.120, 0.442] & $0.155^{\text{ns}}$ & $0.151$ [-0.025, 0.308] \\
 & Sens. & $0.025^{\dagger}$ & $0.221$ [0.053, 0.405] & $0.002^{\ddagger}$ & $0.299$ [0.136, 0.480] & $0.002^{\ddagger}$ & $0.299$ [0.136, 0.480] & $0.008^{\ddagger}$ & $0.257$ [0.090, 0.440] \\
 & Spec. & $0.008^{\ddagger}$ & $-0.231$ [-0.391, -0.099] & $0.079^{\text{ns}}$ & $-0.178$ [-0.330, -0.033] & $0.079^{\text{ns}}$ & $-0.178$ [-0.330, -0.033] & $0.028^{\dagger}$ & $-0.205$ [-0.362, -0.066] \\
 & NSD  & $0.060^{\text{ns}}$ & $0.148$ [-0.020, 0.333] & $0.002^{\ddagger}$ & $0.294$ [0.126, 0.477] & $<0.001^{\|}$ & $0.331$ [0.162, 0.515] & $0.002^{\ddagger}$ & $0.291$ [0.126, 0.475] \\
\cmidrule(lr){1-10}

\multirow{5}{*}{\textbf{HEART}}
 & DSC  & $0.493^{\text{ns}}$ & $-0.026$ [-0.244, 0.192] & $0.485^{\text{ns}}$ & $0.280$ [0.181, 0.378] & $0.493^{\text{ns}}$ & $-0.026$ [-0.244, 0.192] & $0.493^{\text{ns}}$ & $-0.026$ [-0.244, 0.192] \\
 & IoU  & $0.482^{\text{ns}}$ & $-0.032$ [-0.250, 0.188] & $0.496^{\text{ns}}$ & $0.301$ [0.200, 0.402] & $0.482^{\text{ns}}$ & $-0.032$ [-0.250, 0.188] & $0.482^{\text{ns}}$ & $-0.032$ [-0.250, 0.188] \\
 & Sens. & $<0.001^{\|}$ & $0.762$ [0.512, 1.026] & $<0.001^{\|}$ & $0.353$ [0.293, 0.439] & $<0.001^{\|}$ & $0.762$ [0.512, 1.026] & $<0.001^{\|}$ & $0.762$ [0.512, 1.026] \\
 & Spec. & $<0.001^{\|}$ & $-0.643$ [-1.145, -0.466] & $<0.001^{\|}$ & $-0.220$ [-0.524, -0.010] & $<0.001^{\|}$ & $-0.643$ [-1.145, -0.466] & $<0.001^{\|}$ & $-0.643$ [-1.145, -0.466] \\
 & NSD  & $0.103^{\text{ns}}$ & $-0.264$ [-0.472, -0.045] & $0.206^{\text{ns}}$ & $0.158$ [-0.046, 0.400] & $0.005^{\ddagger}$ & $-0.382$ [-0.606, -0.163] & $0.005^{\ddagger}$ & $-0.376$ [-0.596, -0.160] \\
\cmidrule(lr){1-10}

\multirow{5}{*}{\textbf{LUNGS\_RT}}
 & DSC  & $<0.001^{\|}$ & $0.972$ [0.861, 1.109] & $<0.001^{\|}$ & $0.348$ [0.309, 0.416] & $<0.001^{\|}$ & $0.909$ [0.796, 1.058] & $<0.001^{\|}$ & $0.909$ [0.796, 1.058] \\
 & IoU  & $<0.001^{\|}$ & $0.975$ [0.863, 1.114] & $<0.001^{\|}$ & $0.448$ [0.399, 0.531] & $<0.001^{\|}$ & $0.913$ [0.799, 1.064] & $<0.001^{\|}$ & $0.913$ [0.799, 1.064] \\
 & Sens. & $<0.001^{\|}$ & $0.722$ [0.572, 0.871] & $<0.001^{\|}$ & $0.374$ [0.336, 0.449] & $<0.001^{\|}$ & $0.688$ [0.540, 0.832] & $<0.001^{\|}$ & $0.688$ [0.540, 0.832] \\
 & Spec. & $0.003^{\ddagger}$ & $0.183$ [0.050, 0.333] & $<0.001^{\|}$ & $0.313$ [0.211, 0.425] & $0.004^{\ddagger}$ & $0.170$ [0.038, 0.318] & $0.004^{\ddagger}$ & $0.170$ [0.038, 0.318] \\
 & NSD  & $<0.001^{\|}$ & $0.534$ [0.403, 0.677] & $<0.001^{\|}$ & $0.557$ [0.444, 0.672] & $<0.001^{\|}$ & $0.538$ [0.406, 0.682] & $<0.001^{\|}$ & $0.536$ [0.407, 0.680] \\
\cmidrule(lr){1-10}

\multirow{5}{*}{\textbf{LUNGS\_LT}}
 & DSC  & $<0.001^{\|}$ & $1.460$ [1.336, 1.610] & $<0.001^{\|}$ & $0.565$ [0.502, 0.667] & $<0.001^{\|}$ & $1.486$ [1.352, 1.656] & $<0.001^{\|}$ & $1.486$ [1.352, 1.656] \\
 & IoU  & $<0.001^{\|}$ & $1.474$ [1.348, 1.623] & $<0.001^{\|}$ & $0.653$ [0.591, 0.739] & $<0.001^{\|}$ & $1.503$ [1.366, 1.673] & $<0.001^{\|}$ & $1.503$ [1.366, 1.673] \\
 & Sens. & $<0.001^{\|}$ & $1.024$ [0.903, 1.158] & $<0.001^{\|}$ & $0.568$ [0.509, 0.643] & $<0.001^{\|}$ & $1.028$ [0.898, 1.164] & $<0.001^{\|}$ & $1.028$ [0.898, 1.164] \\
 & Spec. & $<0.001^{\|}$ & $0.623$ [0.540, 0.737] & $<0.001^{\|}$ & $0.345$ [0.285, 0.594] & $<0.001^{\|}$ & $0.628$ [0.538, 0.752] & $<0.001^{\|}$ & $0.628$ [0.538, 0.752] \\
 & NSD  & $<0.001^{\|}$ & $0.677$ [0.572, 0.801] & $<0.001^{\|}$ & $0.642$ [0.552, 0.748] & $<0.001^{\|}$ & $0.679$ [0.573, 0.801] & $<0.001^{\|}$ & $0.677$ [0.571, 0.801] \\
\cmidrule(lr){1-10}

\multirow{5}{*}{\textbf{TRACHEA}}
 & DSC  & $<0.001^{\|}$ & $1.166$ [0.761, 1.997] & $0.003^{\ddagger}$ & $0.827$ [0.495, 1.367] & $<0.001^{\|}$ & $1.101$ [0.717, 1.811] & $<0.001^{\|}$ & $1.101$ [0.717, 1.811] \\
 & IoU  & $<0.001^{\|}$ & $1.168$ [0.754, 2.051] & $0.003^{\ddagger}$ & $0.827$ [0.489, 1.387] & $<0.001^{\|}$ & $1.104$ [0.714, 1.856] & $<0.001^{\|}$ & $1.104$ [0.714, 1.856] \\
 & Sens. & $<0.001^{\|}$ & $1.424$ [0.910, 2.378] & $<0.001^{\|}$ & $1.297$ [0.876, 2.347] & $<0.001^{\|}$ & $1.411$ [0.906, 2.384] & $<0.001^{\|}$ & $1.411$ [0.906, 2.384] \\
 & Spec. & $0.600^{\text{ns}}$ & $-0.222$ [-0.634, 0.328] & $0.489^{\text{ns}}$ & $-0.243$ [-0.675, 0.303] & $0.600^{\text{ns}}$ & $-0.227$ [-0.647, 0.322] & $0.600^{\text{ns}}$ & $-0.227$ [-0.647, 0.322] \\
 & NSD  & $0.299^{\text{ns}}$ & $0.375$ [-0.116, 0.747] & $0.299^{\text{ns}}$ & $0.375$ [-0.116, 0.748] & $0.299^{\text{ns}}$ & $0.372$ [-0.116, 0.727] & $0.299^{\text{ns}}$ & $0.362$ [-0.116, 0.697] \\
\cmidrule(lr){1-10}

\multirow{5}{*}{\textbf{BODY}}
 & DSC  & $<0.001^{\|}$ & $0.116$ [0.015, 0.298] & $<0.001^{\|}$ & $0.727$ [0.628, 0.866] & $<0.001^{\|}$ & $0.116$ [0.015, 0.298] & $<0.001^{\|}$ & $0.116$ [0.015, 0.298] \\
 & IoU  & $<0.001^{\|}$ & $0.127$ [0.024, 0.307] & $<0.001^{\|}$ & $0.743$ [0.643, 0.879] & $<0.001^{\|}$ & $0.127$ [0.024, 0.307] & $<0.001^{\|}$ & $0.127$ [0.024, 0.307] \\
 & Sens. & $<0.001^{\|}$ & $-0.003$ [-0.088, 0.114] & $<0.001^{\|}$ & $0.315$ [0.222, 0.421] & $<0.001^{\|}$ & $-0.003$ [-0.088, 0.114] & $<0.001^{\|}$ & $-0.003$ [-0.088, 0.114] \\
 & Spec. & $<0.001^{\|}$ & $0.285$ [0.150, 0.456] & $<0.001^{\|}$ & $0.449$ [0.368, 0.571] & $<0.001^{\|}$ & $0.285$ [0.150, 0.456] & $<0.001^{\|}$ & $0.285$ [0.150, 0.456] \\
 & NSD  & $0.066^{\text{ns}}$ & $-0.021$ [-0.130, 0.070] & $0.066^{\text{ns}}$ & $-0.021$ [-0.130, 0.070] & $0.066^{\text{ns}}$ & $-0.021$ [-0.130, 0.070] & $0.066^{\text{ns}}$ & $-0.021$ [-0.130, 0.070] \\
\cmidrule(lr){1-10}

\multirow{5}{*}{\textbf{GTV}}
 & DSC  & $<0.001^{\|}$ & $0.632$ [0.388, 0.940] & $<0.001^{\|}$ & $0.653$ [0.400, 0.957] & $<0.001^{\|}$ & $0.694$ [0.440, 1.014] & $<0.001^{\|}$ & $0.785$ [0.516, 1.114] \\
 & IoU  & $<0.001^{\|}$ & $0.627$ [0.380, 0.935] & $<0.001^{\|}$ & $0.647$ [0.391, 0.956] & $<0.001^{\|}$ & $0.687$ [0.431, 1.011] & $<0.001^{\|}$ & $0.775$ [0.507, 1.108] \\
 & Sens. & $<0.001^{\|}$ & $0.675$ [0.459, 0.947] & $<0.001^{\|}$ & $0.684$ [0.471, 0.958] & $<0.001^{\|}$ & $0.695$ [0.486, 0.967] & $<0.001^{\|}$ & $0.720$ [0.514, 0.989] \\
 & Spec. & $0.026^{\dagger}$ & $-0.308$ [-0.565, -0.058] & $0.029^{\dagger}$ & $-0.287$ [-0.559, -0.034] & $0.037^{\dagger}$ & $-0.280$ [-0.550, -0.029] & $0.053^{\text{ns}}$ & $-0.241$ [-0.524, 0.020] \\
 & NSD  & $0.853^{\text{ns}}$ & $-0.010$ [-0.261, 0.245] & $<0.001^{\|}$ & $0.369$ [0.124, 0.712] & $0.010^{\ddagger}$ & $0.259$ [0.015, 0.580] & $<0.001^{\|}$ & $0.369$ [0.124, 0.712] \\
\cmidrule(lr){1-10}

\multirow{5}{*}{\textbf{CTV}}
 & DSC  & $0.006^{\ddagger}$ & $0.319$ [0.076, 0.568] & $0.336^{\text{ns}}$ & $0.123$ [-0.108, 0.349] & $0.354^{\text{ns}}$ & $0.123$ [-0.115, 0.353] & $0.006^{\ddagger}$ & $0.319$ [0.076, 0.568] \\
 & IoU  & $0.007^{\ddagger}$ & $0.315$ [0.073, 0.566] & $0.339^{\text{ns}}$ & $0.122$ [-0.108, 0.347] & $0.341^{\text{ns}}$ & $0.122$ [-0.116, 0.350] & $0.007^{\ddagger}$ & $0.315$ [0.073, 0.566] \\
 & Sens. & $<0.001^{\|}$ & $0.940$ [0.664, 1.286] & $<0.001^{\|}$ & $0.814$ [0.553, 1.136] & $<0.001^{\|}$ & $0.812$ [0.552, 1.139] & $<0.001^{\|}$ & $0.940$ [0.664, 1.286] \\
 & Spec. & $<0.001^{\|}$ & $-0.380$ [-0.655, -0.164] & $<0.001^{\|}$ & $-0.420$ [-0.711, -0.202] & $<0.001^{\|}$ & $-0.420$ [-0.708, -0.201] & $<0.001^{\|}$ & $-0.380$ [-0.655, -0.164] \\
 & NSD  & $0.446^{\text{ns}}$ & $0.005$ [-0.212, 0.263] & $0.007^{\ddagger}$ & $0.190$ [-0.038, 0.513] & $0.504^{\text{ns}}$ & $-0.003$ [-0.215, 0.257] & $0.007^{\ddagger}$ & $0.190$ [-0.038, 0.513] \\
\bottomrule
\end{tabular*}

\end{minipage}

\vspace{.5 cm}

\subsection*{S2. Computational Efficiency}

\noindent
\begin{minipage}{\textwidth}
\centering
\small
\captionof{table}{Computational efficiency on an NVIDIA T4 GPU (16~GB VRAM). Lower values are better for all columns.}
\label{tab:compute}
\setlength{\tabcolsep}{4pt}
\renewcommand{\arraystretch}{1.15}
\resizebox{\textwidth}{!}{%
\begin{tabular}{lccccc}
\toprule
\textbf{Model}
 & \textbf{Params (M) $\downarrow$}
 & \textbf{FLOPs (G) $\downarrow$}
 & \textbf{Inf.\ Time (ms/slice) $\downarrow$}
 & \textbf{Mem.\ (MB) $\downarrow$}
 & \textbf{Train (h) $\downarrow$} \\
\midrule
Ours (DAMM-Net++) & 21.7 & 72.6 & 22.4 & 3,128 & 172.4 \\
nnUNet            & 31.2 & 42.1 & 14.2 & 3,452 & 185.6 \\
nnMamba           & 15.55 & 51.6 & 16.3 & 2,612 & 158.2 \\
UNETR++           & 46.7 & 89.4 & 72.5 & 4,896 & 312.8 \\
SwinUNETR         & 62.2 & 105.8 & 76.4 & 5,234 & 348.6 \\
\bottomrule
\end{tabular}%
}
\end{minipage}

\clearpage

\newpage
\subsection*{S3. Uncertainty Calibration and Sparsification Analysis}
\vspace{.5cm}
\noindent
\begin{minipage}{\textwidth}
\centering
\includegraphics[width=1\linewidth,height=1\textheight,keepaspectratio]{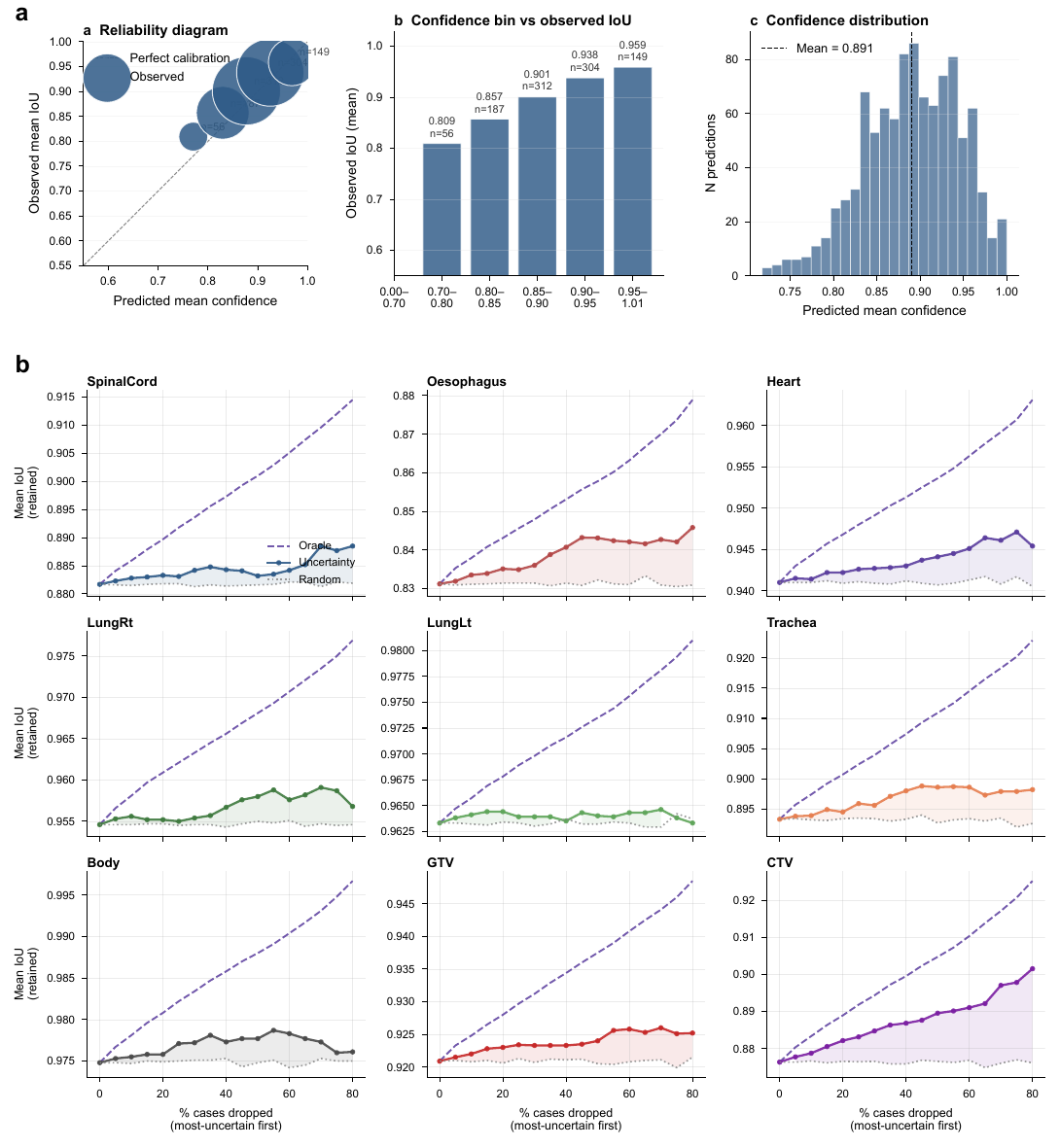}
\captionof{figure}{Uncertainty analysis of \ours{} on the internal test set ($n=2{,}745$ predictions). (a) Confidence calibration: reliability diagram showing predicted confidence versus observed IoU per bin (dashed line indicates perfect calibration), observed IoU by confidence bin with bin counts, and marginal distribution of predicted confidence (dashed line marks the mean). (b) Per-structure sparsification curves: mean IoU on retained set as uncertain cases (highest entropy) are progressively dropped; solid lines show uncertainty ranking, dashed purple shows the oracle upper bound, and dotted grey shows the random baseline. Shaded areas indicate area under the sparsification curve (AUS).}
\label{supfig:uncertainty_analysis}
\end{minipage}

\clearpage

\newpage 
\subsection*{S4. Multi-Planar Prediction Overlays}
\vspace{.5cm}
\noindent
\begin{minipage}{\textwidth}
\centering
\includegraphics[width=1\linewidth,height=1\textheight,keepaspectratio]{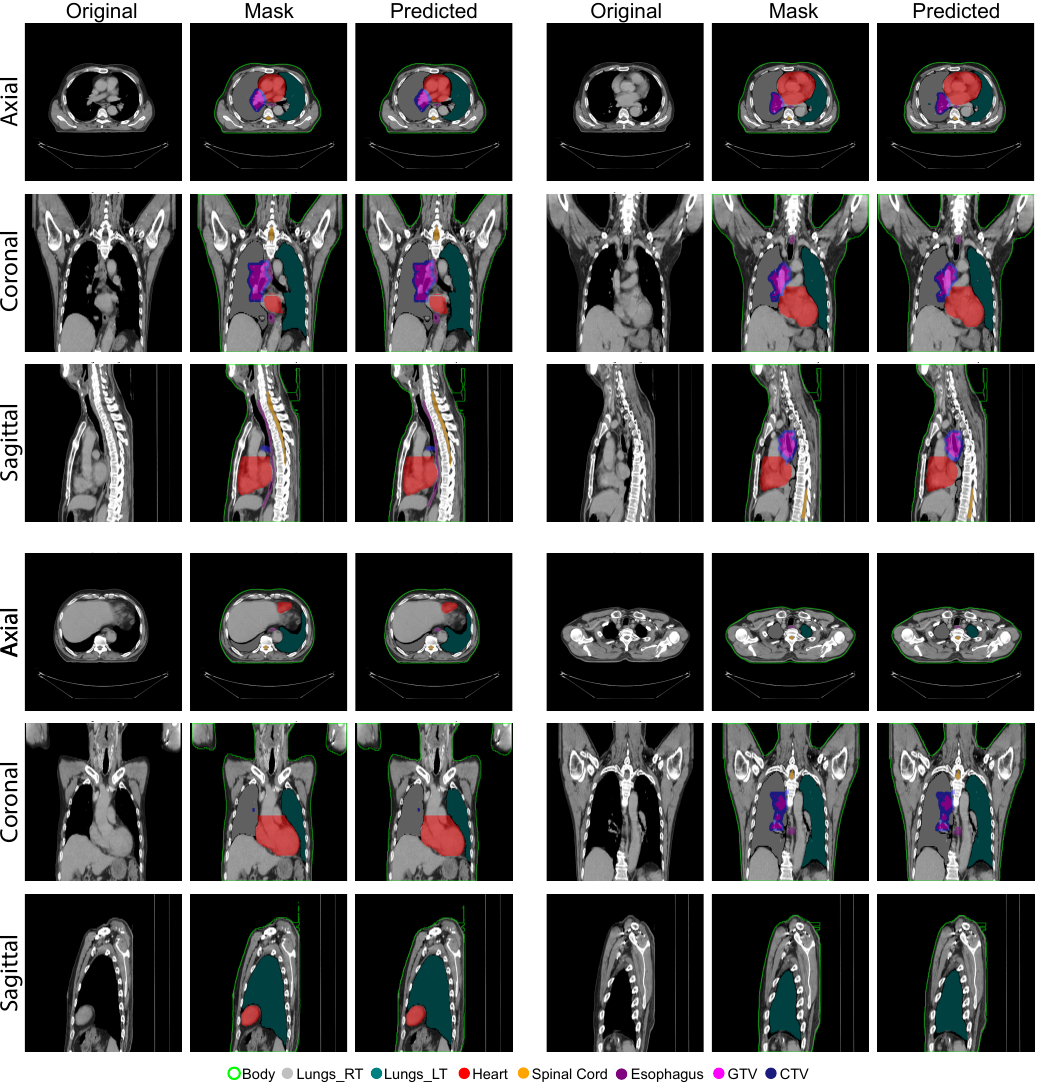}
\captionof{figure}{Through-plane consistency of the proposed model's predictions on a representative thoracic case. Rows show axial (top), coronal (middle), and sagittal (bottom) planes reformatted from the same volumetric prediction. Columns show (1) original CT, (2) ground-truth overlay, and (3) \ours{} prediction overlay. The contours remain spatially coherent across all three orthogonal views, confirming that per-slice predictions translate to geometrically consistent three-dimensional volumes.}
\label{supfig:orthogonal_overlays_detail}
\end{minipage}

\clearpage

\newpage
\subsection*{S5. Additional Ablation Studies}
\vspace{.5cm}
\noindent
\begin{minipage}{\textwidth}
\centering
\small
\captionof{table}{Through-plane memory ablation: Dice, HD95, parameters, latency, and Dice loss relative to selective SSM. Blue and rose mark best and second-best per metric, respectively. The selective SSM achieves the best accuracy-latency trade-off among all alternatives.}
\label{tab:s9_memory_ablation}
\setlength{\tabcolsep}{4pt}
\renewcommand{\arraystretch}{1.15}
\begin{tabular}{lccccc}
\toprule
Memory & Dice$\uparrow$ & HD95$\downarrow$ & Params\,(M)$\downarrow$ & ms/vol$\downarrow$ & $\Delta$Dice \\
\midrule
None (per-slice)          & 0.9382 & 5.31 & \cellcolor{bestblue}$\bm{21.4}$ & \cellcolor{bestblue}$\bm{42}$  & \textcolor{deltadown}{$\downarrow\,0.0163^{**}$} \\
ConvLSTM                  & 0.9451 & 4.63 & 24.9 & 71  & \textcolor{deltadown}{$\downarrow\,0.0094^{**}$} \\
ConvGRU                   & 0.9463 & 4.55 & 23.8 & 66  & \textcolor{deltadown}{$\downarrow\,0.0082^{**}$} \\
Self-attention            & 0.9478 & 4.40 & 26.1 & 88  & \textcolor{deltadown}{$\downarrow\,0.0067^{**}$} \\
Non-selective SSM (S4)    & \cellcolor{secondred}$\bm{0.9491}$ & \cellcolor{secondred}$\bm{4.18}$ & \cellcolor{secondred}$\bm{23.1}$ & \cellcolor{secondred}$\bm{58}$ & \textcolor{deltadown}{$\downarrow\,0.0054^{*}$} \\
\textbf{Selective SSM (ours)} & \cellcolor{bestblue}$\bm{0.9545}$ & \cellcolor{bestblue}$\bm{3.78}$ & 23.6 & 61 & -- \\
\bottomrule
\end{tabular}

\end{minipage}

\vspace{0.5cm}

\noindent
\begin{minipage}{\textwidth}
\centering
\small
\captionof{table}{Internal design of the inter-slice transition branch (top) and Memory-Guided Boundary-Aware decoder (bottom). Each block ablates one design axis; the final row of each block is the full configuration used in DAMM-Net++. Blue and rose mark the best and second-best design within each block, respectively. Multi-scale processing, channel-spatial attention, memory-guided skip attention, and boundary-aware refinement each contribute positively, with the largest gains on surface metrics (NSD, HD95).}
\label{tab:s9_module_design}
\setlength{\tabcolsep}{4pt}
\renewcommand{\arraystretch}{1.15}
\begin{tabular}{lcccc}
\toprule
Variant & Dice$\uparrow$ & NSD$\uparrow$ & HD95$\downarrow$ & ASD$\downarrow$ \\
\midrule
\multicolumn{5}{l}{\emph{Inter-slice transition branch}}\\
Single-scale, no attention        & 0.9487 & 0.9331 & 4.22 & 2.24 \\
Multi-scale, no attention         & 0.9509 & 0.9349 & 4.05 & 2.16 \\
Multi-scale + channel attn.       & \cellcolor{secondred}$\bm{0.9524}$ & \cellcolor{secondred}$\bm{0.9362}$ & \cellcolor{secondred}$\bm{3.95}$ & \cellcolor{secondred}$\bm{2.11}$ \\
\;\;+ spatial gate (full)         & \cellcolor{bestblue}$\bm{0.9545}$ & \cellcolor{bestblue}$\bm{0.9385}$ & \cellcolor{bestblue}$\bm{3.78}$ & \cellcolor{bestblue}$\bm{2.05}$ \\
\midrule
\multicolumn{5}{l}{\emph{Memory-Guided Boundary-Aware decoder}}\\
Plain U-Net decoder               & 0.9463 & 0.9281 & 4.71 & 2.44 \\
+ Memory-guided skip attn.        & 0.9501 & 0.9333 & 4.29 & 2.27 \\
+ Boundary-aware refinement       & \cellcolor{secondred}$\bm{0.9527}$ & \cellcolor{secondred}$\bm{0.9371}$ & \cellcolor{secondred}$\bm{3.94}$ & \cellcolor{secondred}$\bm{2.10}$ \\
\;\;+ deep boundary sup. (full)   & \cellcolor{bestblue}$\bm{0.9545}$ & \cellcolor{bestblue}$\bm{0.9385}$ & \cellcolor{bestblue}$\bm{3.78}$ & \cellcolor{bestblue}$\bm{2.05}$ \\
\bottomrule
\end{tabular}

\end{minipage}

\vspace{0.5cm}

\noindent
\begin{minipage}{\textwidth}
\centering
\small
\captionof{table}{Hyperparameter sensitivity (Dice). Bold marks the default used throughout. Blue marks the best value within each sweep. The chosen defaults ($d_{\text{state}}=16$, sequence length $T=5$, three transition scales) achieve the highest Dice across all settings, with clear degradation observed for extremely small or large values. $T=1$ corresponds to single-slice (2D) baseline; $T=\text{All}$ processes the full volumetric stack.}
\label{tab:s9_hyperparameters}
\setlength{\tabcolsep}{6pt}
\renewcommand{\arraystretch}{1.15}
\begin{tabular}{lccccccc}
\toprule
SSM state $d_{\text{state}}$ & 4 & 8 & \textbf{16} & 24 & 32 & 48 & 64 \\
\midrule
Dice & 0.9401 & 0.9482 & \cellcolor{bestblue}$\bm{0.9545}$ & 0.9540 & 0.9532 & 0.9501 & 0.9478 \\
\midrule
Sequence length $T$ & 1 & 3 & \textbf{5} & 7 & 9 & 11 & All \\
\midrule
Dice & 0.9047 & 0.9411 & \cellcolor{bestblue}$\bm{0.9545}$ & 0.9538 & 0.9519 & 0.9498 & 0.9384 \\
\midrule
Transition scales & 1 & 2 & \textbf{3} & 4 & 5 & 6 & — \\
\midrule
Dice & 0.9487 & 0.9518 & \cellcolor{bestblue}$\bm{0.9545}$ & 0.9539 & 0.9521 & 0.9494 & — \\
\bottomrule
\end{tabular}

\end{minipage}

\newpage
\subsection*{S6. Consolidated Reader Study Analysis}
\vspace{1cm}
\noindent
\begin{minipage}{\textwidth}
\centering
\includegraphics[width=\linewidth,height=1\textheight,keepaspectratio]{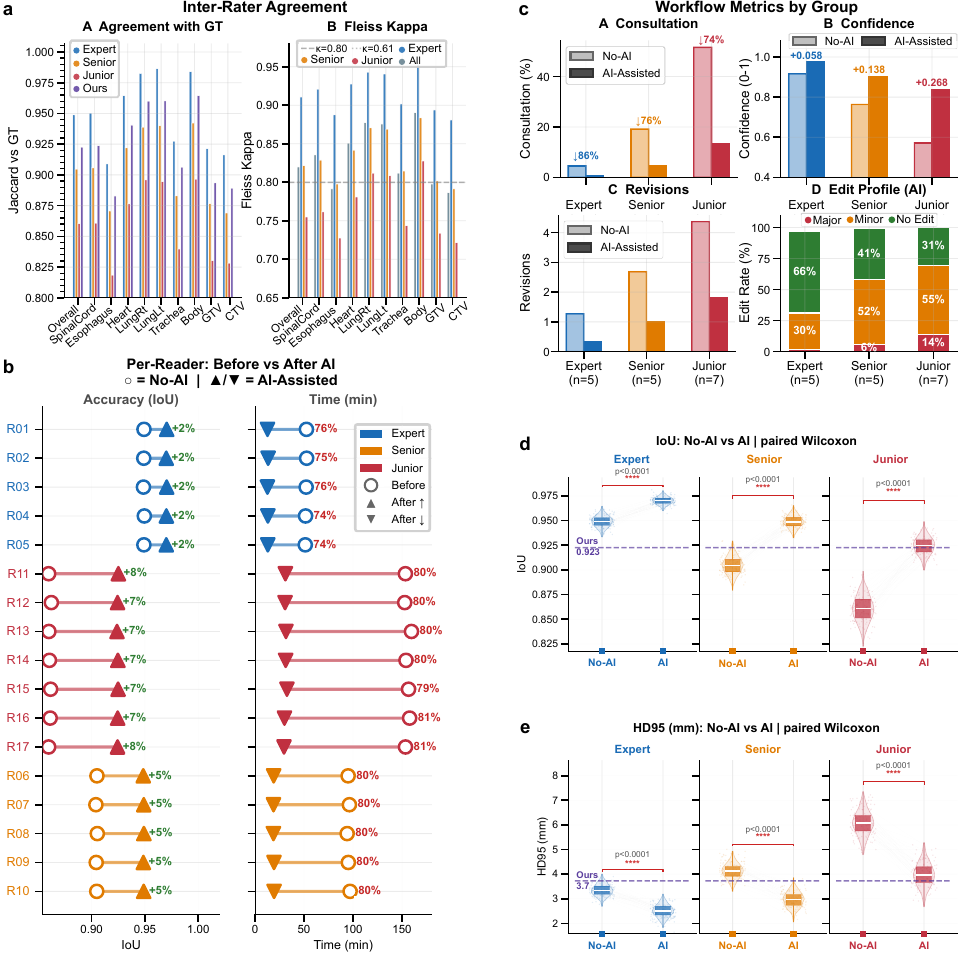}
\captionof{figure}{Consolidated reader study analysis across 17 radiation oncologists (5 experts, 5 seniors, 7 juniors) on 305 test-set cases under No-AI and AI-Assisted conditions. (a) Inter-rater and reference-standard agreement: structure-wise Jaccard against the expert-panel consensus contour by reader tier and \ours{} (left), and Fleiss' $\kappa$ by structure and tier (right); dashed lines mark substantial ($\kappa=0.61$) and excellent ($\kappa=0.80$) agreement thresholds. \ours{} matched the expert range on every OAR and outperformed juniors on GTV and CTV. (b) Workflow metrics by tier: (A) consultation rate; (B) reader confidence; (C) contour revisions; (D) edit-type profile. AI assistance reduced the junior consultation rate from $51.6\%$ to $13.5\%$ and reached a $65.7\%$ no-edit rate for experts. (c) Per-reader clinical impact: mean IoU (left) and contouring time (right), with open circles for No-AI and filled markers for AI-Assisted. All 17 readers improved on both metrics. (d) Paired within-reader IoU comparison by tier; dashed line: unedited model reference (IoU $=0.9226$). Wilcoxon tests rejected the null in every group ($p<0.0001$), with Hedges' $g=+4.30$ (Expert), $+5.59$ (Senior), and $+5.58$ (Junior). (e) Paired within-reader HD95 comparison (mm; lower is better); dashed line: model mean HD95 ($3.73$~mm). HD95 improved significantly in every group ($p<0.0001$), with Hedges' $g=+3.39$ (Expert), $+3.71$ (Senior), and $+4.31$ (Junior).}
\label{supfig:reader_study_consolidated}
\end{minipage}

\clearpage

\newpage
\subsection*{S7. Per-Class Reader-Tier Comparison}
\vspace{.5cm}
\noindent
\begin{minipage}{\textwidth}
\centering
\includegraphics[width=.90\linewidth,height=.62\textheight,keepaspectratio]{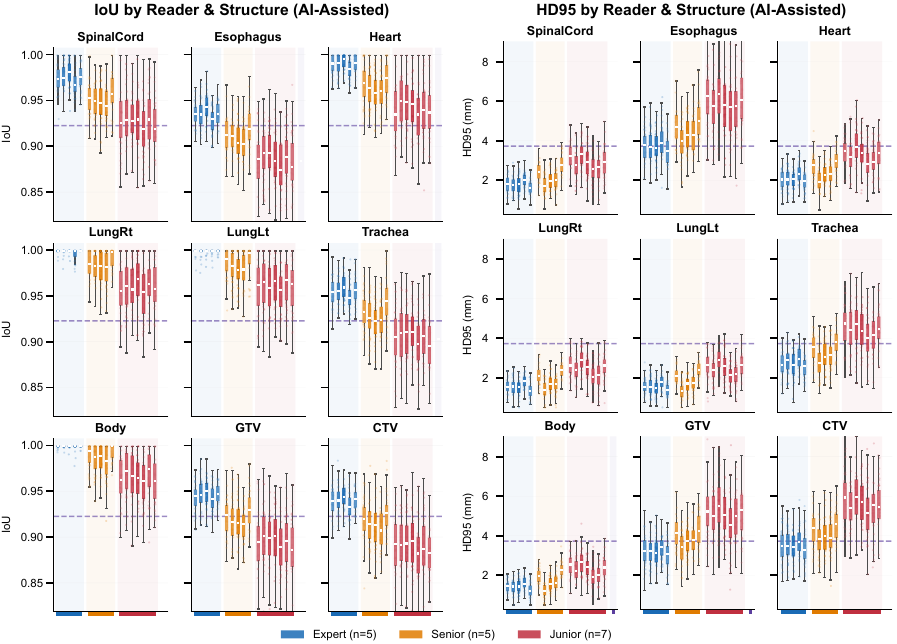}
\captionof{figure}{Per-class reader-tier comparison across 17 radiation oncologists (5 experts, 5 seniors, 7 juniors) and the unedited \ours{} model. Each panel shows nine structure subpanels with 18 tiny boxes (5 experts, 5 seniors, 7 juniors, \ours{}); colored tier bands run beneath the bottom-row subpanels, and the dashed line marks the \ours{} reference. (a) Per-class IoU distributions: boxes cluster tightly within the Expert tier and spread widest in the Junior tier; \ours{} reaches Junior-tier level on all structures. (b) Per-class HD95 distributions (mm; lower is better): within-tier spread is largest for Juniors on low-contrast targets (esophagus, GTV, CTV), and \ours{} delivers the largest HD95 improvements over Juniors on CTV, GTV, and trachea.}
\label{supfig:reader_study_perclass}
\end{minipage}

\vspace{1cm}

\noindent
\begin{minipage}{\textwidth}
\centering
\small
\captionof{table}{Reader-study primary and secondary endpoint estimates from the linear mixed-effects model (Section~\ref{sec:clinical_evaluation}), fit on $10{,}370$ case-level observations. Reported: coefficient $\beta$, standard error, 95\% CI, standardized effect, raw $p$, and Holm-adjusted $p$. Junior is the reference group for AI $\times$ Group interactions. Random effects: reader ICC $=0.003$, case ICC $=0.002$ (Time); all ICCs $\leq 0.004$. Period-effect estimates were $\leq 0.5\%$ of AI effect magnitude on all endpoints.}
\label{tab:supp_mixedmodel}
\setlength{\tabcolsep}{4pt}
\renewcommand{\arraystretch}{1.15}
\begin{tabular}{llcccccc}
\toprule
\textbf{Endpoint} & \textbf{Contrast}
   & $\bm{\beta}$ & \textbf{SE}
   & \textbf{95\% CI}
   & \textbf{Std.\ effect}
   & $\bm{p}$
   & $\bm{p_{\text{Holm}}}$ \\
\midrule
\multirow{3}{*}{Time (min)}
   & AI (Junior ref.)      & $-124.39$ & $0.87$ & $[-126.10,\,-122.68]$ & $-2.86$ & $<\!0.0001$ & $<\!0.0001$ \\
   & AI $\times$ Expert    & $+85.38$  & $0.65$ & $[+84.10,\,+86.66]$   & $+1.96$ & $<\!0.0001$ & $<\!0.0001$ \\
   & AI $\times$ Senior    & $+48.05$  & $0.65$ & $[+46.77,\,+49.32]$   & $+1.11$ & $<\!0.0001$ & $<\!0.0001$ \\
\midrule
\multirow{3}{*}{IoU}
   & AI (Junior ref.)      & $+0.0638$ & $0.0004$ & $[+0.0631,\,+0.0646]$ & $+1.66$ & $<\!0.0001$ & $<\!0.0001$ \\
   & AI $\times$ Expert    & $-0.0426$ & $0.0003$ & $[-0.0431,\,-0.0421]$ & $-1.11$ & $<\!0.0001$ & $<\!0.0001$ \\
   & AI $\times$ Senior    & $-0.0198$ & $0.0003$ & $[-0.0204,\,-0.0193]$ & $-0.52$ & $<\!0.0001$ & $<\!0.0001$ \\
\midrule
\multirow{3}{*}{HD95 (mm)}
   & AI (Junior ref.)      & $-2.095$  & $0.023$ & $[-2.140,\,-2.050]$   & $-1.62$ & $<\!0.0001$ & $<\!0.0001$ \\
   & AI $\times$ Expert    & $+1.255$  & $0.017$ & $[+1.221,\,+1.290]$   & $+0.97$ & $<\!0.0001$ & $<\!0.0001$ \\
   & AI $\times$ Senior    & $+0.938$  & $0.017$ & $[+0.904,\,+0.972]$   & $+0.72$ & $<\!0.0001$ & $<\!0.0001$ \\
\midrule
Consultation (log-odds)
   & AI (Junior ref.)      & $-1.952$  & $0.077$ & $[-2.10,\,-1.80]$      & N/A    & $<\!0.0001$ & $<\!0.0001$ \\
\bottomrule
\end{tabular}

\end{minipage}

\newpage
\subsection*{S8. External Validation Analysis}
\vspace{.5cm}
\noindent
\begin{minipage}{\textwidth}
\centering
\includegraphics[width=\linewidth,height=0.62\textheight,keepaspectratio]{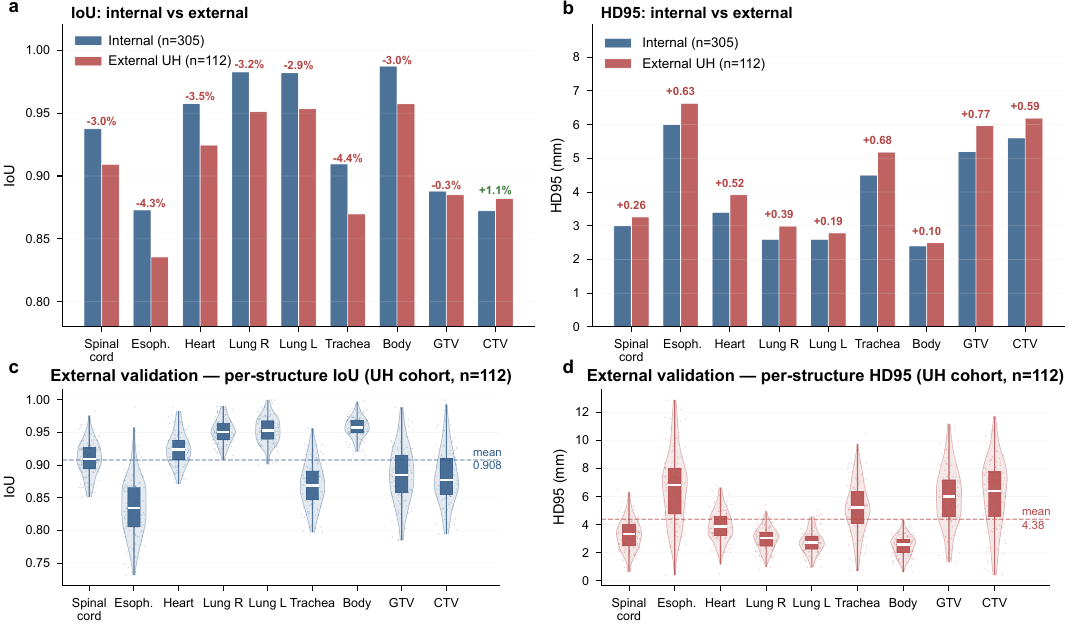}
\captionof{figure}{External validation on the UH cohort ($n=112$). (a) Internal versus external IoU across the nine structures, showing per-structure gaps between the multi-institutional internal test cohort ($n=305$) and the external UH cohort. Mean IoU decreased from 0.932 (internal) to 0.908 (external), a relative gap of $-2.64\%$, with the largest drop on the trachea ($-4.4\%$) and a small gain on the CTV ($+1.1\%$). No structure exceeded a $-5\%$ drop. (b) Internal versus external HD95, showing the corresponding boundary-error gaps (mm; lower is better). All structures showed modest increases in HD95, with the largest absolute increase on the GTV ($+0.77$~mm) and the smallest on the body ($+0.10$~mm). (c) Per-structure case-level IoU distributions on the external UH cohort. Box: IQR; central white tick: median; whiskers: $1.5\times$IQR fence; dots: individual cases. The gradient follows the expected structure-difficulty pattern, with body and lungs near the top and soft-tissue targets (esophagus, GTV, CTV) at the bottom. (d) Per-structure case-level HD95 distributions (mm; lower is better) on the same cohort. All $n=112$ cases are shown per structure.}
\label{supfig:external_validation}
\end{minipage}

\vspace{1cm}

\noindent
\begin{minipage}{\textwidth}
\centering
\small
\captionof{table}{Confidence calibration on the external UHL cohort ($n=1{,}008$ predictions). Predictions binned by mean confidence; per-bin absolute calibration error is $|\text{bin mean confidence} - \text{observed IoU}|$. Weighted absolute calibration error: $0.020$.}
\label{tab:ext_calibration}
\setlength{\tabcolsep}{4pt}
\renewcommand{\arraystretch}{1.15}
\begin{tabular}{lcccc}
\toprule
Confidence bin & $n$ & Bin mean confidence & Observed IoU (mean) & Absolute cal.\ error \\
\midrule
0.00--0.70 &   0 & N/A   & N/A    & N/A \\
0.70--0.80 &  56 & 0.771 & 0.8094 & 0.0387 \\
0.80--0.85 & 187 & 0.830 & 0.8573 & 0.0274 \\
0.85--0.90 & 312 & 0.877 & 0.9012 & 0.0239 \\
0.90--0.95 & 304 & 0.925 & 0.9382 & 0.0129 \\
0.95--1.01 & 149 & 0.969 & 0.9590 & \cellcolor{bestblue}$\bm{0.0098}$ \\
\bottomrule
\end{tabular}
\end{minipage}
\clearpage

\newpage
\subsection*{S9. Volumetric 3D Reconstruction}
\vspace{.5cm}
\noindent
\begin{minipage}{\textwidth}
\centering
\captionsetup{skip=6pt}
\includegraphics[width=.58\linewidth,height=0.62\textheight,keepaspectratio]{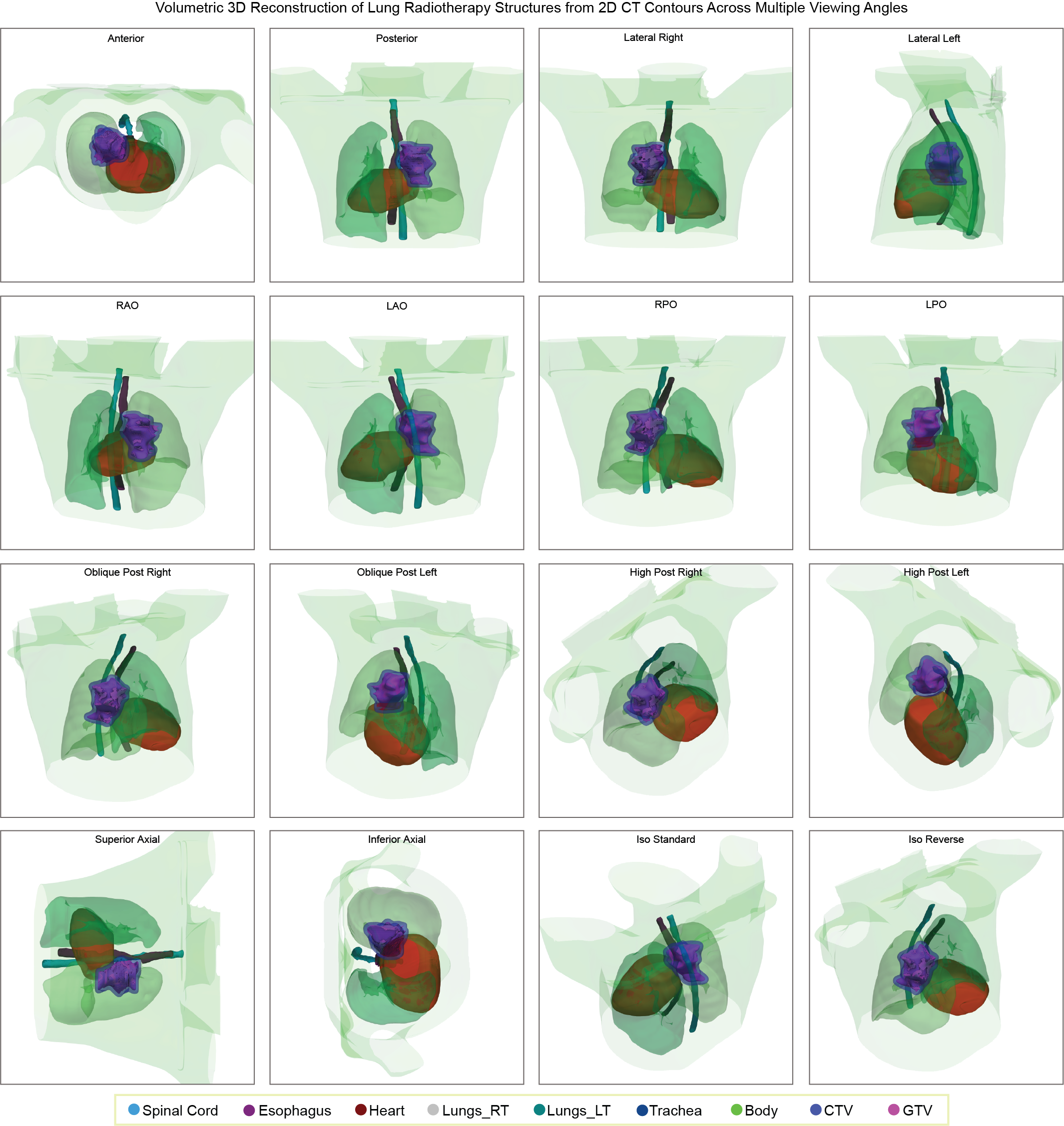}
\captionof{figure}{Volumetric 3D reconstruction of nine structures across sixteen clinical viewing angles. Color coding: Spinal Cord (cyan), Esophagus (purple), Heart (red), Right Lung (light green), Left Lung (dark green), Trachea (dark teal), Body (transparent), CTV (navy), GTV (dark maroon).}
\label{fig:3d_recon}
\end{minipage}

\par\addvspace{12pt}
\subsection*{S10. Review Application Interface}

\noindent
\begin{minipage}{\textwidth}
\centering
\captionsetup{skip=6pt}
\includegraphics[width=.9\linewidth,height=0.62\textheight,keepaspectratio]{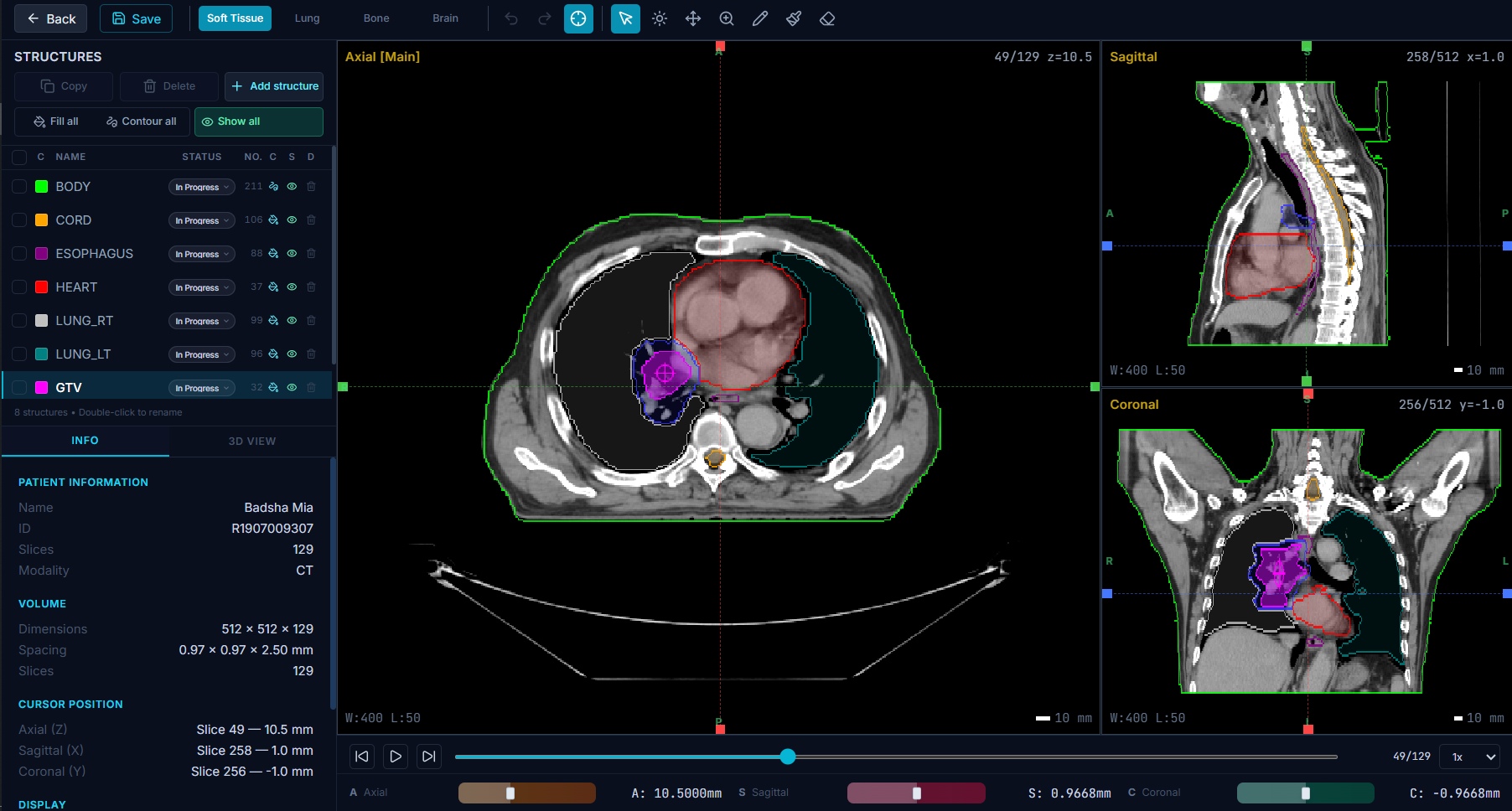}
\captionof{figure}{Screenshot of the \ourmethod{} review application. Three-panel orthogonal display (left) shows CT in axial, coronal, and sagittal planes with predicted contours as color-coded overlays; left panel provides per-structure visibility and opacity controls.}
\label{fig:webapp}
\end{minipage}

\onecolumn
\subsection*{S11. Training Hyperparameters}
\noindent
\begin{minipage}{\textwidth}
\centering
\small
\captionof{table}{Training hyperparameters used for \ours{}.}
\label{tab:s10_hyperparams}
\setlength{\tabcolsep}{4pt}
\renewcommand{\arraystretch}{1.05}
\begin{tabular*}{\linewidth}{@{\extracolsep{\fill}}lp{.60\linewidth}@{}}
\toprule
\textbf{Parameter} & \textbf{Value} \\
\midrule
Optimizer & AdamW \\
Initial learning rate & $10^{-4}$ \\
Weight decay & $10^{-5}$ \\
Betas & (0.9, 0.999) \\
Epsilon & $10^{-8}$ \\
Learning rate schedule & CosineAnnealingWarmRestarts \\
$T_0$ & 50 \\
$T_{mult}$ & 2 \\
$\eta_{min}$ & $10^{-6}$ \\
Batch size & 2 sequences \\
Sequence length & 5 slices \\
Effective batch size & 10 slices \\
Epochs & 200 \\
Early stopping patience & 80 \\
Gradient clipping & max\_norm = 1.0 \\
Input resolution & 512 $\times$ 512 \\
Loss weights & $\boldsymbol{\lambda} = \{1.0, 0.5, 0.2, 0.15, 0.2, 0.1, 0.1, 0.1\}$ \\
\bottomrule
\end{tabular*}
\end{minipage}

\par\vspace{10pt}

\subsection*{S12. DICOM-to-NIfTI Conversion and Multi-Planar Slice Extraction}
\vspace{.5cm}
\noindent
\begin{minipage}{\textwidth}
\centering
\footnotesize
\captionsetup[figure]{labelformat=unnumberedalgorithm,labelsep=colon,justification=raggedright,singlelinecheck=false}
\captionof{figure}{DICOM-to-NIfTI conversion and multi-planar slice extraction pipeline.}
\label{alg:dicom_pipeline}
\begin{algorithmic}
\Require DICOM CT series and RTSTRUCT file for each patient
\Ensure CT NIfTI volume, multi-class segmentation NIfTI, axial/coronal/sagittal PNG slices
\State \textbf{Step 1: Load data}
\State Load DICOM CT volume and RTSTRUCT into 3D Slicer
\State Extract patient identifier from DICOM tag \texttt{(0010,0020)}
\State Export CT volume as \texttt{.nii.gz}
\State \textbf{Step 2: Extract valid structures}
\For{target structure $s$}
    \State Extract binary labelmap
    \If{labelmap contains no non-zero voxels}
        \State Skip segment $s$
    \EndIf
\EndFor
\State Assign canonical labels
\State \textbf{Step 3: Build multi-label mask}
\State Initialize $\mathbf{L} \leftarrow \mathbf{0}$
\For{structure in priority order}
    \State $\mathbf{L}[\text{mask}_{s} > 0] \leftarrow \ell(s)$
\EndFor
\State Export $\mathbf{L}$ as compressed multi-class NIfTI
\State \textbf{Step 4: CT windowing and normalization}
\State $W_c \leftarrow 50$, $W_w \leftarrow 400$
\State $\mathbf{H}_{\text{clip}} \leftarrow \text{clip}(\mathbf{H},\; -150,\; 250)$
\State $\mathbf{I} \leftarrow \left\lfloor \frac{\mathbf{H}_{\text{clip}} + 150}{400} \times 255 \right\rfloor$ cast to \texttt{uint8}
\State \textbf{Step 5: Extract and save multi-planar slices}
\For{axial slice $z$}
    \State Extract $\mathbf{I}_{z}$; replicate to RGB; save as PNG
    \State Extract $\mathbf{L}_{z}$; apply per-class palette; save as PNG
\EndFor
\For{coronal slice $y$}
    \State Extract and save corresponding CT and segmentation slices
\EndFor
\For{sagittal slice $x$}
    \State Extract and save corresponding CT and segmentation slices
\EndFor
\State \Return CT NIfTI, multi-class segmentation NIfTI, axial/coronal/sagittal PNG slices
\end{algorithmic}
\end{minipage}

\end{document}